\documentclass[notitlepage,10pt,aps,prd,twocolumn,amsmath,amssymb,floatfix,superscriptaddress,nofootinbib,preprintnumbers]{revtex4-1}

\usepackage{amsfonts,amssymb,amsmath,graphicx,color,bm,epsfig}
\usepackage[normalem]{ulem}
\usepackage{multirow}
\usepackage{enumerate}
\usepackage[utf8]{inputenc}
\usepackage{hyperref}
\usepackage{braket}
\usepackage{subfigure}
\usepackage{float}
\usepackage[dvipsnames]{xcolor}
\usepackage{soul}
\usepackage{rotating}

\hypersetup{
    colorlinks=true,
    linkcolor=red,
    citecolor=blue,
}

\allowdisplaybreaks

\begin{document}
\title{Reconstructing Gravity on Cosmological Scales}
\author{Marco Raveri}
\affiliation{Kavli Institute for Cosmological Physics, Department of Astronomy \& Astrophysics, Enrico Fermi Institute, The University of Chicago, Chicago, IL 60637, USA}
%
%%%%%%%%%%%%%%%%%%%%%%%%%%%%%%%%%%%%%%%%%%%%%%%%%%%%%%%%%%%%%%%%%%%%%%%%
\begin{abstract}
We present the data-driven reconstruction of gravitational theories and Dark Energy models on cosmological scales.
We showcase the power of present cosmological probes at constraining these models and quantify the knowledge of their properties that can be acquired through state of the art data.
This reconstruction exploits the power of the Effective Field Theory approach to Dark Energy and Modified Gravity phenomenology, which compresses the freedom in defining such models into a finite set of functions that can be reconstructed across cosmic times using cosmological data.
We consider several model classes described within this framework and thoroughly discuss their phenomenology and data implications. 
We find that some models can alleviate the present discrepancy in the determination of the Hubble constant as inferred from the cosmic microwave background and as directly measured.
This results in a statistically significant preference for the reconstructed theories over the standard cosmological model.
\end{abstract}
%%%%%%%%%%%%%%%%%%%%%%%%%%%%%%%%%%%%%%%%%%%%%%%%%%%%%%%%%%%%%%%%%%%%%%%%
\maketitle
%%%%%%%%%%%%%%%%%%%%%%%%%%%%%%%%%%%%%%%%%%%%%%%%%%%%%%%%%%%%%%%%%%%%%%%%
%
%%%%%%%%%%%%%%%%%%%%%%%%%%%%%%%%%%%%%%%%%%%%%%%%%%%%%%%%%%%%%%%%%%%%%%%%
\section{Introduction} \label{Sec:Intro}
%%%%%%%%%%%%%%%%%%%%%%%%%%%%%%%%%%%%%%%%%%%%%%%%%%%%%%%%%%%%%%%%%%%%%%%%

The physical mechanism behind cosmic acceleration~\cite{Riess:1998cb,Perlmutter:1998np} still poses a challenge to our theoretical understanding.
Its phenomenology, in turn, is the target of current and future observational efforts. These plan to exploit the increasing precision of measurements of the cosmic microwave background (CMB) and the large scale structures (LSS) of our Universe to characterize its properties.
These observations can be used to study and constrain models where cosmic acceleration is achieved through a new dark component, Dark Energy (DE), 
or with modifications to the laws of gravity on large scales (for reviews, see Refs.~\cite{Silvestri:2009hh,Clifton:2011jh,Joyce:2014kja}).

Both CMB measurements and LSS data have proven to be extremely powerful in pursuing this program.
Existing observations can be explained to large extent in simple terms, with the $\Lambda$CDM model in which cosmic acceleration is sourced by a cosmological constant.
Current measurements already constrain deviations from this scenario~\cite{Ade:2015rim,Joudaki:2016kym,Aghanim:2018eyx,Alam:2016hwk,Abbott:2018xao} while the next generation of probes, such as Euclid~\cite{Laureijs:2011gra}, LSST~\cite{Abell:2009aa}, and CMB-S4~\cite{Abazajian:2016yjj} are expected to largely exceed their performances~\cite{Zhao:2008bn,Zhao:2009fn,Hojjati:2011xd,Asaba:2013mxj} and characterize the properties of both DE and modified gravity (MG) models to unprecedented precision.

At the same time, hints of discrepancies between different cosmological probes, within the $\Lambda$CDM model have been found.
The expansion rate of the Universe as derived from the CMB differs from distance ladder measurements~\cite{Riess:2018byc}.
Data from LSS surveys and the CMB show different pictures of how cosmological structures grew over time~\cite{Abbott:2017wau}.
Even though these discrepancies were barely noteworthy in the past, their statistical significance continues to steadily increase~\cite{Raveri:2018wln} and might point toward the fact that we are close to a radical, paradigm shifting discovery.

To fully and efficiently exploit the power of cosmological observations at testing DE and MG scenarios the Effective Field Theory (EFT) of DE and MG~\cite{Gubitosi:2012hu,Bloomfield:2012ff} was created to be able to describe the cosmological phenomenology of both families of models with the same language, creating a model independent theoretical tool that describes most of the models that have been studied so far.
This formulation eliminates possible redundancies in the definition of DE/MG models by retaining only the properties that are relevant for cosmology.
As such the EFT of DE compresses the freedom that we have in defining a DE/MG theory, at the cosmological level, in a limited set of functions of time.

In this paper we present their first complete late-time reconstruction from the publicly available cosmological data.
We reconstruct several families of models, described within the EFT of DE, starting from simple Quintessence ones up to the full family of Horndeski models. We thoroughly comment on their cosmological implications and data constraints.

We show that this approach greatly outperforms approaches consisting of simple parametrizations and is able to efficiently harness the data constraining power.
We quantify how much can be learned, through the data, with present cosmological observations, combining probes of the expansion history of the universe with measurements of the CMB and LSS.

We comment on the role that these models play in explaining some of the discrepancies between different cosmological data sets.
In particular we find that some of them alleviate the present tension between the CMB and local measurements of the Hubble constant. This solution is also consistent with supernovae and baryon acoustic oscillations measurements.
Because of the close connection between the EFT treatment that we use and specific physical models this resolution can be achieved within well defined DE or MG models that satisfy theoretical consistency conditions.

This paper is organized as follows:
in Sec.~\ref{Sec:EFT} we briefly review our EFT approach;
in Sec.~\ref{Sec:ReconstructionMethod} we elaborate on the technique used to constrain the EFT operators with the cosmological data sets that we detail in Sec.~\ref{Sec:Method};
in Sec.~\ref{Sec:Results} we describe, model by model, our findings that we summarize in Sec.~\ref{Sec:Conclusions}.

%%%%%%%%%%%%%%%%%%%%%%%%%%%%%%%%%%%%%%%%%%%%%%%%%%%%%%%%%%%%%%%%%%%%%%%%
\section{Parametrized Dark Energy and Gravity} \label{Sec:EFT}
%%%%%%%%%%%%%%%%%%%%%%%%%%%%%%%%%%%%%%%%%%%%%%%%%%%%%%%%%%%%%%%%%%%%%%%%
%
We focus on a broad class of theories that includes the majority of DE and MG models, with a viable cosmology, that have been studied in literature.
This consists of models with an extra scalar propagating degree of freedom with respect to General Relativity that are also known as scalar-tensor theories.
We consider models with second order equations of motion which, in their most general form, belong to the model class known as Horndeski gravity~\cite{Horndeski:1974wa,Deffayet:2011gz,Kobayashi:2011nu}.
Our purpose is that of reconstructing the large-scale cosmological behavior of models within this class as described in the EFT of DE approach in the form of its lowest order operators.

With respect to the original Horndeski formulation that describes different DE/MG models with several free functions of the scalar field and its kinetic energy, the EFT approach, gets rid of possible redundancies by compressing the freedom in defining them in a limited set of functions of time only~\cite{Piazza:2013coa,Gleyzes:2013ooa}.
This makes the EFT language especially suited to efficiently explore the observational implications of such models from an EFT perspective~\cite{Raveri:2014cka,Hu:2014sea,Hu:2015rva,Bellini:2015xja,Raveri:2017qvt,Kreisch:2017uet,Peirone:2017ywi,Mancini:2018qtb,Reischke:2018ooh,Espejo:2018hxa,Noller:2018wyv,SpurioMancini:2019rxy} while at the same time retaining the possibility to test specific models once they are mapped to the EFT of DE~\cite{Frusciante:2015maa,Hu:2016zrh,Peirone:2016wca,Peirone:2017vcq,Benevento:2018xcu,Raveri:2018ddi}.

We do not include in this work higher order operators that, although present in the EFT expansion, would generally correspond to beyond Horndeski models and Lorentz violating theories~\cite{Kase:2014cwa,Gleyzes:2014dya,Frusciante:2015maa}. 
The investigation of their general cosmological implications is left for future work.
We also only consider models that satisfy the weak equivalence principle with all matter species universally and minimally coupled to gravity.
Notice that we do not explicitly consider vector-tensor and tensor-tensor theories for two reasons: their lowest order EFT operators for the scalar mode can be matched to the EFT operators that we consider~\cite{Lagos:2017hdr}; there is no measured cosmological observable that would significantly constrain the extra vector and tensor modes.

% What is the EFT basis that we consider:
Several equivalent formulations of the EFT of DE exist in literature~\cite{Piazza:2013coa,Gleyzes:2013ooa,Bellini:2014fua} and we choose, following~\cite{Bellini:2014fua,Raveri:2017qvt}, to work with a basis that is specified by the following five functions of time:
\begin{itemize}
\item $\Lambda(t)$: a time dependent cosmological constant, whose time dependence is usually present in all models of DE beyond $\Lambda$CDM;
\item $M_P^2(t)$: a time dependent ``bare" Planck mass, usually generated by a conformal coupling between the scalar field and gravity;
\item $\alpha_K(t)$: the kinetic energy term in the scalar field Lagrangian that quantifies the independent dynamics of the scalar field;
\item $\alpha_B(t)$: the kinetic mixing between the scalar field and gravity which is commonly found in MG models;
\item $\alpha_T(t)$: the excess speed of gravitational waves and usually non-zero when non-linear derivative couplings of the scalar field to the metric are included in the Lagrangian.
\end{itemize}
We refer to these functions as the EFT functions and we comment on the details of this parametrization in App.~\ref{App:EFT}.
The $\Lambda$CDM model is recovered when: $\Lambda$ is time independent and set to the cosmological constant value $\Lambda(t)=\Lambda$; the Planck mass is time independent and set to the value measured in the solar system $M_P^2(t)=M_P^2$; $\alpha_K(t)=\alpha_B(t)=\alpha_T(t)=0$.
Notice that in practice we consider dimensionless variations of the EFT functions and thus we will work with $\Delta\Lambda/\Lambda$ and $\Delta M_P^2/M_P^2$.

Given the tight relation between the EFT parametrization and specific gravitational models a given choice of the EFT functions has to satisfy requirements of physical viability and in particular should be free from ghost and gradient instabilities in the scalar and tensor sectors at all times~\cite{DeFelice:2016ucp,Frusciante:2018vht}.
This translates in hard bounds in the parameter space of the given EFT model~\cite{Raveri:2014cka,Peirone:2017lgi}.
In this respect we highlight that the specific parametrization that we employ was found in~\cite{Raveri:2017qvt,Peirone:2017ywi,Espejo:2018hxa} to have good stability properties resulting in a wide parameter space that can then be efficiently sampled.

Notice that the fact that a given behavior of the EFT functions satisfies viability constraints ensures the existence of at least one physically viable model in the considered class, as discussed in~\cite{Kennedy:2018gtx}.

% What are the previous constraints:
When considering the terms present in the EFT expansion we do not enforce that the speed of gravitational waves should be equal to the speed of light today, following the multi-messenger detection of the binary neutron star merger event GW170817 and GRB170817A~\cite{Monitor:2017mdv} as the energy scales at which the event was measured are close to the cut-off scale at which cosmological EFT actions become invalid~\cite{deRham:2018red}.
We highlight that both future gravitational waves measurements at lower frequencies~\cite{deRham:2018red} as well as the detection of the CMB B-mode spectrum~\cite{Amendola:2014wma,Raveri:2014eea} might be used to constrain the speed of gravitational waves within the regime of validity of cosmological EFTs.

For the same reason we do not consider solar system constraints and constraints on the Planck mass. These would require the non-linear completion of the EFT describing screening mechanisms (see~\cite{Joyce:2014kja} for a review).

We constrain directly the Planck mass and not only its time dependence because a fixed redefinition of its value, with respect to the solar system value, cannot be reabsorbed in a redefinition of densities.
In particular we are not free to redefine the energy density of radiation that is measured, in a non-gravitational way, through measurements of the temperature of the CMB.
Following this example, we do not redefine the energy density of all matter species, so that the corresponding cosmological parameters maintain the physical meaning of the energy density that would be measured, in a non-gravitational way, by an observer today.

% What are the models that we consider:
In the following we consider several families of models:
\begin{itemize}
\item {\it Quintessence} (Q) models, obtained by considering only the EFT function $\{\Lambda\}$ and fixing the time dependence of $\alpha_K$ as in~\cite{Bellini:2014fua}.
This includes all models where the scalar field is minimally coupled to gravity and with a standard kinetic term (see~\cite{Tsujikawa:2013fta} for a review);
\item {\it K-essence} (K) models, obtained by considering the EFT functions $\{\Lambda,\alpha_K\}$.
This includes models where the scalar field is minimally coupled to gravity but has a non-standard kinetic term~\cite{ArmendarizPicon:1999rj,ArmendarizPicon:2000ah};
\item {\it Kinetic Gravity Braiding} (KGB) models, obtained by considering the EFT functions $\{\Lambda,\alpha_B,\alpha_K\}$ as defined in~\cite{Deffayet:2010qz};
\item {\it Generalized Brans-Dicke} (GBD) models, obtained by considering the EFT functions $\{\Lambda,M_P\}$ and fixing the time dependence of $\alpha_K$ and $\alpha_B$ as in~\cite{Bellini:2014fua}.
This includes all models where the scalar field has a standard kinetic term but is not minimally coupled to gravity.
Special models in this family include Jordan Brans-Dicke~\cite{Brans:1961sx}, $f(R)$~\cite{Hu:2007nk}, chameleons~\cite{Khoury:2003rn} and symmetron models~\cite{Hinterbichler:2010es};
\item {\it Scalar Horndeski} (SH) models, obtained by considering the EFT functions $\{\Lambda,M_P,\alpha_K,\alpha_B\}$ while keeping $\alpha_T=0$.
In this model the speed of gravitational waves is the same as the speed of light at all times.
The SH class includes all previous classes, and allows for non-canonical forms of the kinetic term for the scalar field but without higher derivative couplings;
\item {\it Full Horndeski} (FH) models, obtained by considering all the EFT functions $\{\Lambda,M_P,\alpha_K,\alpha_B,\alpha_T\}$.
This is the most general model that we consider and includes all previous classes.
\end{itemize}
Notice that we consider all these different classes of models, even thought they can all be viewed as sub-spaces of FH, because in the most general model they are a set of measure zero.
As such they would not be explored when a data measure is defined.

% comment on cosmology
When considering these different models we focus on their implications for late time cosmology.
For this reason we are going to restrict to low to mildly high redshifts, $z\in[0,9]$, and this sets the observational imprints that we can observe.
In general all considered models will modify the expansion rate of the universe and this will influence distance determinations.
In addition they will generally influence the growth of perturbations in different ways, depending on the model.
Some models will only result in an overall rescaling of the growth of perturbations while others will modify it in a scale dependent way.
Since we restrict to late times, early times physics is unaltered. This includes possible effects at recombination, discussed in~\cite{Lin:2018nxe}. The structure of the primary peaks in the CMB, in both temperature and polarization, is then unchanged while the CMB spectra might be shifted as a result of late times modification of the expansion history.
Notice that since recombination physics is unchanged with respect to the standard scenario the BAO position remains a standard ruler~\cite{Lin:2018nxe}.
Most of the changes to the CMB spectra are going to be driven by changes in the CMB lensing potential, as a result of the modified growth, and the late-time ISW effect.
The modified growth of perturbations is directly probed by measurements of the clustering and lensing of galaxies.

% calculation of derived quantities:
We will project the results of the reconstruction on some phenomenological quantities to characterize the cosmological behavior of the different models.
We first consider the effective DE equation of state, defined as:
\begin{align} \label{Eq:wDEeff}
w_{\rm DE} \equiv \frac{P_{\rm DE}}{\rho_{\rm DE}}
\equiv \frac{-M_{P0}^2(3H^2+2\dot{H}) -P_m}{3M_{P0}^2H^2-\rho_m} \,,
\end{align}
where $M_{P0}^2$ denotes the present day value of the Planck mass as measured in the solar system, $P_m$ and $\rho_m$ the total energy density and pressure of all the matter components, $H$ the Hubble factor and the over-dot represents derivatives with respect to cosmic time.
Notice that as soon as we consider models where the scalar field is not minimally coupled to gravity the effective DE density can easily transition from positive to negative and make $w_{\rm DE}$ ill-defined.
This is not generally a pathology of the background expansion history but is a failure mode of the effective definition.
In such cases we will resort to the effective total equation of state defined as: 
\begin{align} \label{Eq:weffTot}
w_{\rm tot} \equiv -\frac{3H^2+2\dot{H}}{3H^2} \,,
\end{align}
that does not suffer from these pathologies and is regular at all times.
In addition we compare the expansion history of a given model to the expansion history that we would have had in the $\Lambda$CDM model with the same cosmological parameters,
$\Delta H / H_{\Lambda \rm CDM} \equiv (H-H_{\Lambda \rm CDM})/H_{\Lambda \rm CDM}$.

%%%%%%%%%%%%%%%%%%%%%%%%%%%%%%%%%%%%%%%%%%%%%%%%%%%%%%%%%%%%%%%%%%%%%%%%
\section{Reconstructing functions of time} \label{Sec:ReconstructionMethod}
%%%%%%%%%%%%%%%%%%%%%%%%%%%%%%%%%%%%%%%%%%%%%%%%%%%%%%%%%%%%%%%%%%%%%%%%
%
% what is the goal in general terms:
We seek to reconstruct the time dependence of the EFT functions, for different models, as we discussed in the previous section.
This reconstruction approach has already been successfully applied to phenomenological DE/MG properties~\cite{Huterer:2002hy,Mortonson:2008qy,Mortonson:2009hk,Zhao:2012aw,Vanderveld:2012ec,Wang:2015wga,Zhao:2017cud,Miranda:2017mnw,Li:2018nlh,Wang:2018fng,Li:2018dcf} and we extend it to EFT models reconstruction.

With respect to the reconstruction of phenomenological properties of DE/MG the reconstruction of the EFT operators has both advantages and disadvantages. 
The fact that in the latter approach the reconstructed models have to satisfy constraints on their physical viability makes the parameter space of the reconstruction harder to explore. At the same time this means that the selected models do correspond to healthy physical theories, which is not guaranteed by phenomenological reconstructions.
On the other hand the former approach, in its most general form, is able to extract all interesting and statistically significant features in the data while the EFT reconstruction would only single out the ones that correspond to single scalar field DE/MG models.

% what are the challenges in general terms and how do we solve them
Reconstructing a generic function of time from the data is challenging, mainly for two reasons. 
In this section we review these challenges and present how we address them, mostly following the spirit of~\cite{Crittenden:2011aa}.
We highlight that we assume that all the EFT functions are continuous.
In addition their derivatives, up to the third, directly enter the dynamical equations and have to be continuous as well.

% interpolation and infinite dimensional:
The first challenging aspect of the reconstruction is the fact that the space of all functions on an interval is infinite dimensional.
Since the amount of measurements that we have available is finite, the space where the data would define a probability measure is finite dimensional.
Even if we were to define a measure on the space of functions it cannot be put in any meaningful correspondence to the measure defined over the space of data.

This problem is addressed by interpolation techniques (see for example~\cite{isaacson1994analysis}).
These establish optimal methods to represent the infinite dimensional space of functions with an appropriate finite dimensional space given by the value of the function at a finite set of mesh points.

For this reason we are going to represent all the EFT functions as a collection of values on a fixed time grid and interpolate these values with a piece-wise fifth order spline.
This choice has two advantages. Spline interpolation has close to optimal rate of convergence, over the space of functions, as we increase the number of mesh points. 
If a suitably high order for the spline is chosen then not only the function is going to be continuous but also its derivatives will be continuous and converge to the derivatives of the function that they are representing.
The order of the spline is chosen to ensure that all derivatives of the EFT functions that enter in the dynamical equations are continuous.

Notice that these considerations immediately exclude some common choices.
We do not use Taylor expansions to represent the EFT functions because their rate of convergence over the space of functions is well known to be extremely slow.
We also exclude piece-wise continuous interpolation (i.e.~binning) as its derivatives would not converge to the derivatives of the target function.

% The second problem is the truncation order, solved by smoothness priors and searches in the RKHS space
The second challenge in reconstructing functions of time is that, after choosing a function basis, we need to choose a truncation order for the function expansion.
In absence of a clear criterion to stop as we increase the order of the expansion, the model is going to progressively fit better the data until the order matches the number of data points. 
Then all higher order contributions are going to be unconstrained and the corresponding solution will over-fit the data and the model will lack predictivity.
The solution to this problem is to restrict the reconstruction to the space of smooth functions (see for example~\cite{3569}) for an appropriate smoothing kernel.

Following~\cite{Crittenden:2011aa,Zhao:2012aw,Zhao:2017cud,Wang:2018fng} we work with the Gaussian smoothing kernel defined by the CPZ correlation~\cite{Crittenden:2011aa} that only depends on the correlation time of the process.
An estimate of the correlation time for the models that we consider is discussed in~\cite{Raveri:2017qvt,Espejo:2018hxa} and so we choose the correlation time of the reconstruction prior to be $0.3$ in scale factor units.
Notice that this choice is not particularly important as it effectively enforces the smoothness of the reconstruction as a low pass filter, disfavoring fast variations but allowing functions smoother on scales larger than the correlation time to fit the data unpenalized.
Our focus is to perform the reconstruction at late times and thus we choose the scale factor range to be $a\in [0.1,1]$.
At times earlier than the beginning of the reconstruction we force the EFT functions to match their GR limit.
This means that we have three correlation lengths in the reconstruction range, effectively penalizing functions that exhibit more than six extrema.
Results in this last correlation length can be influenced by our requirement to match $\Lambda$CDM cosmology at high redshift and therefore we will read off results at later times, from either the redshift of DM/DE equality, $z_{\rm eq}$ or the redshift of the beginning of cosmic acceleration, $z_{\rm acc}$. Both redshifts are about two correlation lengths away from the $\Lambda$CDM limit so possible effects induced by the boundary will have had time to decay.

The choice of reconstructing the sub-space of smooth functions automatically sets the truncation order to use in the functional expansion.
As soon as we have enough mesh points per correlation time the reconstruction is converged as further features, described by additional mesh points, are going to be increasingly disfavored by the prior.
Still we need to prove that convergence is achieved and to do so we study three cases where we have five, ten and fifteen equispaced mesh points in $a\in [0.1,1]$ per EFT function, corresponding to roughly one, three and five mesh points per correlation length.
In addition to qualitative checks on the end results we also test that increasing number of mesh points are sampling from the same posterior distribution.
In practice we consider samples from the posterior of the reconstruction for two different grids, with ten and fifteen points respectively.
We can merge the samples of the posterior with fifteen mesh points with the samples from the posterior with ten points over-sampled to the fifteen point grid.
Then we use standard statistical tools to check the convergence to the target distribution and in particular we can compute the Gelman and Rubin~\cite{Gelman:1992zz,An98stephenbrooks} $R$ test on the joint samples and verify that a target value of $R-1<0.1$ is achieved. Notice that this is an additional test and that we require all models to satisfy $R-1<0.01$ on their own.
For the joint test we lower the threshold because the convergence of the ten mesh points chain is degraded by over-sampling.

The last detail to make the reconstruction complete over the space of smooth functions concerns the bias toward the assumed mean when using smoothness priors.
To overcome this we apply the smoothness Gaussian priors, as in~\cite{Crittenden:2011aa}, between a given sample of function values at the mesh points and its smoothed version, obtained with a Gaussian kernel with a variance that matches the correlation length of the prior, evaluated at the mesh points.
This means that we effectively penalize only functional modes that do not belong to the space of smooth functions.
This approach performs better than bin average, discussed in~\cite{Crittenden:2011aa}, as it remains invariant under changes of the time grid.

% how do we set the overall strength of the prior
To set the overall strength of the smoothness prior, for each model, we perform parameter estimation with all relevant EFT functions set to a smoothed step function that transitions away from GR at the beginning of the reconstruction range and is constant after that.
For a given number of reconstruction mesh points the overall strength of the prior is chosen so that the variance of the most constrained prior mode is two orders of magnitude weaker than that of constant mode. 

% prior range:
In addition we set the a hard prior range to be:
$\Lambda(a)\in [-2,2]$; $M_p(a)\in [-1,1]$; $\alpha_K(a)\in [-10,10]$; $\alpha_B(a)\in [-10,10]$; $\alpha_T(a)\in [-10,10]$.
These are approximately four orders of magnitude weaker than the constraint on the constant mode and ensure that the value range of the reconstruction is wide.
The only exception is $\alpha_K$, whose constant mode is unconstrained, and we arbitrarily fix it to be one order of magnitude larger than the other EFT functions. In this case the strength of the correlation prior is set to be one order of magnitude tighter than the hard prior bound.

% relation with PCA:
Overall we believe that the reconstruction strategy that we discussed performs better than principal components (PC) based reconstructions~\cite{Mortonson:2008qy,Mortonson:2009hk,Vanderveld:2012ec,Miranda:2017mnw}.
The PC basis is obtained from forecasts around a given cosmology, usually the best fitting $\Lambda$CDM one, relies on the Gaussianity of the parameter posterior and when truncated is not guaranteed to be able to describe statistically significant deviations from it.
Both problems are not present in our reconstruction scheme that is not biased toward a fiducial cosmology.
In addition, excluding some PCs from the fit means that we assume perfect knowledge on them and fix their variance to zero.
This will influence the overall variance of the reconstruction artificially reducing it.

% what are the quantities that we compute after the fact
After performing the reconstruction we use several statistical techniques to extract interesting information from it.

% how do we quantify how much we have learned: the number of effective parameters
We quantify the knowledge learned through the reconstruction, following~\cite{Raveri:2018wln}, by computing the Gaussian approximation of the number of effective parameters that are supported by the data:
\begin{align} \label{Eq:NumberEffParams}
N_{\rm eff} \equiv N -{\rm tr}[ \mathcal{C}_\Pi^{-1}\mathcal{C}_p ] \,,
\end{align}
where $N$ is the total nominal number of parameters of the model, $\mathcal{C}_\Pi$ is the prior parameter covariance and $\mathcal{C}_p$ is the posterior parameter covariance.
Notice that the prior covariance for the EFT reconstruction does not only contain the effect of the smoothness prior but also the effect of the stability conditions that effectively modify the shape (and hence the covariance) of the parameter space. For this reason we perform a Markov Chain Monte Carlo (MCMC) prior-only parameter estimation run for each model.
%

% how do we quantify the performances of the model: best fit and evidence
To quantify the data performances of a given reconstruction we compute the evidence ratio, $\Delta \log_{10}\mathcal{E}\equiv \log_{10}\mathcal{E}_{\Lambda{\rm CDM}}-\log_{10}\mathcal{E}_{\rm EFT}$, with respect to the $\Lambda$CDM model in two ways: from the reconstruction samples, as in~\cite{Heavens:2017afc}; with a Gaussian approximation, as in~\cite{Raveri:2018wln}.
We are going to interpret the results, as commonly done in literature, on Jeffreys' scale~\cite{Jeffreys61,kassr95}
for which $3:1$ odds for one of the model is ``substantial'' evidence, $10:1$ is ``strong'', $30:1$ is ``very strong'', $100:1$ is ``decisive''.

We compute and highlight the Gaussian approximation of the Occam's razor factor present in the evidence calculation due to the additional prior volume in the reconstruction.
For a single model this is defined as:
\begin{align} \label{Eq:OccamsRazor}
\log_{10}\mathcal{O} = \frac{1}{2}\log_{10}\left(|\mathcal{C}_{\Pi}^{-1}\mathcal{C}_{p}| \right) \,,
\end{align}
where $|\cdot|$ denotes the determinant.
The Gaussian approximation of the evidence ratio is then just the sum of the maximum posterior, $\mathcal{P}_{\rm max}$, change and the Occam's razor factor
$\Delta \log_{10}\mathcal{E} = \Delta \log_{10}\mathcal{P}_{\rm max} +\Delta \log_{10}\mathcal{O}$.
The evidence ratio usually heavily penalizes the reconstruction because of the much wider parameter space.
This is, however, not entirely fair to the reconstruction which is not meant to have good performances on its own but should be viewed as an aid for model builders to develop theories that perform well with the data and achieve similar results with just a handful of parameters.
For this reason we also provide the signal to noise ratio (SNR) of the reconstruction as the difference in data $\chi^2$ with respect to the $\Lambda$CDM model.
We compute the SNR significance by thinking that the difference in data $\chi^2$ could be achieved with one extra parameter and then convert the probability value to the effective number of standard deviations defined by $n_{\sigma}^{\rm eff}(P) \equiv \sqrt{2}{\rm Erf}^{-1} (P)$, where ${\rm Erf}^{-1}$ is the inverse error function.

% how do we tell what information is coming from the prior and data: KL decomposition, marginal KL and best fitting sub-spaces
To visualize the reconstruction modes that the data is constraining we use the Karhunen-Lo\`eve (KL) decomposition of the prior ($\Pi$) and posterior ($p$) covariances.
To do so we solve the generalized eigenvalue problem to find the KL modes, $\phi^{a}$, of the two covariances:
\begin{align}
\sum_{\nu} \mathcal{C}_{\Pi}^{\mu\nu} \, \phi_{\nu}^{\,\,a} = \lambda^{a} \sum_{\nu} \mathcal{C}_{p}^{\mu\nu} \, \phi_{\nu}^{\,\,a} \,.
\end{align}
The eigenmodes are defined to be orthonormal in the $\mathcal{C}_{p}$ metric
\begin{equation}
\sum_{\mu\nu} \phi_{\mu}^{\,\,a}\mathcal{C}_{p}^{\mu\nu}  \phi_{\nu}^{\,\,b} = \delta^{ab},
\end{equation}
and since they are orthogonal in the $\mathcal{C}_{\Pi}$ metric, but with variance $\lambda^a \delta^{ab}$, the KL basis provides linear combinations of the parameters that are mutually independent and ordered by the improvement in the variance of the posterior over the prior.
Notice that a given $\lambda^a$ quantifies the ratio of the prior to posterior KL mode variance, that we indicate as $\sqrt{\lambda^a} \equiv \sigma_{\Pi}/\sigma_{\rm KL}^a$.
We also project the posterior over the best constrained KL modes to visualize the best fitting reconstruction subspace.
This is an useful projection because it removes the information that is contaminated by the prior and clearly shows what the data is constraining.
The constraints are usually tighter because the KL mode filtering represents what we know about this particular EFT function rather than what we do not know.
The constraints are then indicative of the minimal space where the model would provide the better fit to the data.
Further variations that are not contained in this space would, by definition, be unconstrained by the data and hence cannot improve the performances of the model.
The KL mode filtering is easily done, after computing the KL modes by projecting the posterior, sample by sample, eliminating the poorly constrained modes and re-projecting the samples on the original basis.
We choose to filter out modes that have $\sigma_{KL}>3 \sigma_\Pi$.

%%%%%%%%%%%%%%%%%%%%%%%%%%%%%%%%%%%%%%%%%%%%%%%%%%%%%%%%%%%%%%%%%%%%%%%%
\section{Data sets and tools} \label{Sec:Method}
%%%%%%%%%%%%%%%%%%%%%%%%%%%%%%%%%%%%%%%%%%%%%%%%%%%%%%%%%%%%%%%%%%%%%%%%
%
% discussion of the data sets:
To perform the EFT reconstruction we employ several data sets.
Since we need to harvest all the constraining power available to be able to efficiently perform the reconstruction we always combine all the data sets that we discuss.

We use the measurements of the CMB temperature and polarization power spectra at small angular scales from the {\it Planck} satellite~\cite{Ade:2015xua, Aghanim:2015xee} supplemented by the large scale {\it Planck} $TEB$ measurements that mainly constrain the optical depth $\tau$. 
We add the {\it Planck} 2015 full-sky lensing potential power spectrum~\cite{Ade:2015zua} in the multipole range $40\leq \ell \leq 400$. 
At smaller angular scales CMB lensing is strongly influenced by the non-linear evolution of dark matter perturbations and we thus exclude multipoles above $\ell=400$.
In addition we add the CMB temperature spectrum measurements at small angular scales from the South Pole Telescope~\cite{2014ApJ...782...74H}.
We exclude all multipoles $\ell>2500$ as they are influenced by the non-linear evolution of the CMB lensing potential.
We refer to the union of these data sets as the CMB data set.

We employ the measurements of the galaxy weak lensing shear correlation function as provided by the Canada-France-Hawaii Telescope Lensing Survey (CFHTLenS)~\cite{Heymans:2013fya,Joudaki:2016mvz}.
We applied ultra-conservative cuts, that make CFHTLenS data insensitive to the modeling of non-linear scales, as in~\cite{Joudaki:2016mvz}.
We call this data set the WL one.

We further include BAO measurements from BOSS DR12~\cite{Alam:2016hwk}, SDSS Main Galaxy Sample~\cite{Ross:2014qpa} and 6dFGS~\cite{Beutler:2011hx}.
From these data sets we exclude redshift-space distortion measurements as they are obtained by assuming that the linear growth of matter perturbations is scale independent while in general DE/MG models the growth is scale dependent.
We indicate this data set as the BAO one.

We consider the Pantheon Supernovae sample~\cite{Scolnic:2017caz} and refer to it as the SN data set.
We use the local determination of the Hubble constant from~\cite{Riess:2018byc} and indicate this measurement as H0.

Notice that we pay a particular attention to the removal of data points that could be influenced by the non-linear evolution of cosmological perturbations and that are beyond the reach of the EFT description that we use.

% tension and expected performances:
Some of the data sets that we use show discrepancies between each other within the $\Lambda$CDM model, as discussed in~\cite{Raveri:2018wln}.
Based on these results we would expect that: fully solving the tension between the CMB and the H0 data set would lead to a decrease in $\chi^2$ of about $12$; solving the discrepancy between CMB and WL measurements would lead to a $\Delta \chi^2$ of about $4$; solving the disagreement between the CMB temperature spectrum and the CMB lensing potential reconstruction would give approximately $\Delta \chi^2=6$.
All these discrepancies combined would result, if resolved, in $\Delta \chi^2=22$. 
In this light we later comment on the performance of DE/MG models to provide an explanation of these tensions.

% discussion of the codes:
To produce cosmological predictions and compare them to data we use the EFTCAMB and EFTCosmoMC codes~\cite{Hu:2013twa,Raveri:2014cka,Hu:2014oga}, modifications, respectively, to the CAMB~\cite{Lewis:1999bs} and CosmoMC~\cite{Lewis:2002ah} codes implementing the EFT of DE.
The cosmological predictions of the EFTCAMB code were validated with several other codes in~\cite{Bellini:2017avd}.
All the tools necessary to perform the EFT reconstruction will be made publicly available in the next release of the EFTCAMB code.

% priors:
In addition to the reconstruction priors discussed in the previous section we use standard priors on the base cosmological parameters (see e.g.~\cite{Raveri:2018wln}).
We fix the sum of neutrino masses to the minimal value~\cite{Long:2017dru}.
We also include all the recommended parameters and priors describing systematic effects in the data sets.

\begin{figure}[tbp]
\centering
\includegraphics[width=\columnwidth]{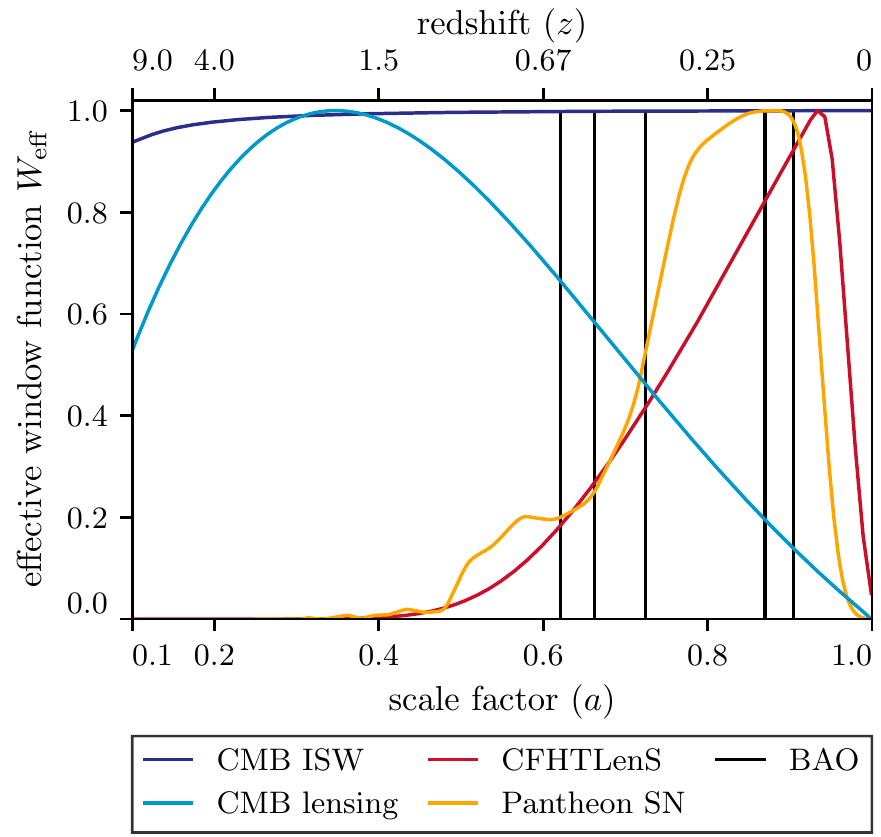}
\caption{ \label{fig:binningspace} 
The effective window function, $W_{\rm eff}$, for the different data sets that we consider, normalized to unity at maximum, over the time range of the EFT reconstruction.
Different colors represent different probes, as shown in legend.
Notice that this is not representative of the overall constraining power of a single data set but only of the times at which a data set is contributing its constraints.
}
\end{figure} 
%

% discussion of how the binning strategy influences or inter plays with the data
%
We now discuss how the reconstruction strategy, and in particular our choice for the correlation time, interplays with the available data sets.
In Fig.~\ref{fig:binningspace} we show the effective window function, $W_{\rm eff}$, normalized to one at the peak, for the different cosmological probes that we use:
the late-time ISW effective window function is defined as $W_{\rm eff}^{\rm ISW} \equiv e^{-\kappa}$ where $\kappa$ is the total optical depth~\cite{Seljak:1996is};
the CMB lensing effective window function is defined as in~\cite{Lewis:2006fu};
the WL effective window function is given by the lensing efficiency~\cite{Kilbinger:2014cea} and we show the total efficiency while the parameter estimation pipeline uses the full tomographic WL data;
the SN effective window function is given by the number of SN in a scale factor bin $dN/da$;
the BAO effective window function shows the different BAO samples effective redshifts.
Notice that the window functions are only illustrative of which times a given dataset constrains cosmology and DE/MG, and in particular does not represent the relative constraining power of different probes.

From Fig.~\ref{fig:binningspace} we can see that a scale factor grid is the natural choice from a data perspective, allowing the full range of the reconstruction to be populated by measurements.
We can also see that the ISW effect is sensitive over the whole reconstruction range, constraining integral modes.
The other data sets naturally divide the reconstruction range in three pieces, corresponding to the three correlation lengths that we have in the reconstruction.
The first is probed by CMB lensing, the second by high-redshift SN, BAO and part of the WL measurements, the low redshift end is probed by SN, BAO and WL.

The fact that one data set would occupy a correlation length, limits the sensitivity of the reconstruction to possible internal, redshift dependent, residual systematic effects in the data.
The overlap between the tails of the different distributions, and given that we have more than one data set in one correlation length,
ensures that the reconstruction is less sensitive to the possibility that one data set has residual overall unaccounted systematic effects.

%%%%%%%%%%%%%%%%%%%%%%%%%%%%%%%%%%%%%%%%%%%%%%%%%%%%%%%%%%%%%%%%%%%%%%%%
\section{Results} \label{Sec:Results}
%%%%%%%%%%%%%%%%%%%%%%%%%%%%%%%%%%%%%%%%%%%%%%%%%%%%%%%%%%%%%%%%%%%%%%%%
%
In this section we present the results of the EFT reconstruction.
In Sec.~\ref{Sec:GeneralResults} we present a series of general results while in each of the following sub-sections we present the results for different families of models: 
in Sec.~\ref{Sec:Quintessence} we discuss the Quintessence (Q) reconstruction;
in Sec.~\ref{Sec:KGB} we discuss the Kinetic Gravity Braiding (KGB) results;
in Sec.~\ref{Sec:GBD} we show the reconstruction of the Generalized Brans-Dicke (GBD) class of models;
in Sec.~\ref{Sec:SH} we show the results of the Scalar Horndeski (SH) reconstruction;
in Sec.~\ref{Sec:FH} we present the results of the Full Horndeski (FH) reconstruction.

%%%%%%%%%%%%%%%%%%%%%%%%%%%%%%%%%%%%%%%%%%%%%%%%%%%%%%%%%%%%%%%%%%%%%%%%
\subsection{General Results} \label{Sec:GeneralResults}
%%%%%%%%%%%%%%%%%%%%%%%%%%%%%%%%%%%%%%%%%%%%%%%%%%%%%%%%%%%%%%%%%%%%%%%%
%
% Number of effective parameters
We start by discussing the constraining power of present cosmological probes over different families of models and the overall knowledge that can be gained with EFT reconstructions.
This is quantified by the number of extra effective parameters, $\Delta N_{\rm eff}$ as in Eq.~(\ref{Eq:NumberEffParams}), that are being constrained, relative to the base $\Lambda$CDM parameters, and the results are shown in Tab.~\ref{Table:ReconstructionNeff}.
We also show there the marginalized $N_{\rm eff}$ for each of the EFT functions, $\{N_{\Lambda},N_{\alpha_K},N_{M_P},N_{\alpha_B},N_{\alpha_T}\}$, obtained by marginalizing the prior and posterior covariances over the other parameters, to identify the functional modes that are better constrained by the data.
\begin{table}[!ht]
\setlength{\tabcolsep}{11pt}
\centering
\begin{tabular}{@{}llllll@{}}
\toprule
model & $\Delta N_{\rm eff}$ & $\{N_{\Lambda},N_{\alpha_K},N_{M_P},N_{\alpha_B},N_{\alpha_T}\}$ \\
\toprule
Q   & $4.2$ & $\{4.2,-,-,-,-\}$ \\
\colrule
K   & $4.2$ & $\{4.2,0.0,-,-,-\}$ \\
\colrule
KGB & $12.0$ & $\{2.4,0.0,-,8.9,-\}$ \\
\colrule
GBD & $14.5$ & $\{2.0,0.0,11.9,-,-\}$ \\
\colrule
SH & $19.8$ & $\{1.5,0.0,10.9,7.2,-\}$ \\
\colrule
FH & $28.3$ & $\{1.7,0.0,10.3,6.8,9.6\}$ \\
\botrule
\end{tabular}
\caption{ \label{Table:ReconstructionNeff}
The total number and the marginalized number of extra effective parameters that are constrained by the data for each model and for the reconstruction of each of the EFT functions that we consider.
Model acronyms are defined in Sec.~\ref{Sec:EFT}.
}
\end{table}

As we can see from Tab.~\ref{Table:ReconstructionNeff} the constraining power of present cosmological probes is remarkably high, over the whole family of Horndeski models.
As we increase the complexity of the theory, adding EFT functions to the reconstruction, the data is able to express greater constraining power.
In the following sections we comment, model by model, on the physical effects that allow this.

The number of effective parameters constrained by the data in different models, as shown in Tab.~\ref{Table:ReconstructionNeff}, can be compared to other methods of constraining these models by simple phenomenological parametrizations of DE/MG properties or the EFT functions.
These results clearly show how EFT reconstruction outperforms simpler approaches.
These are by far not guaranteed to allow the data to show all their constraining power and the results in Tab.~\ref{Table:ReconstructionNeff} show how much these could be missing.
In particular for simple dynamical DE models, parametrizing the equation of state of DE with the CPL~\cite{Chevallier:2000qy,Linder:2002et} functional form, comparing the results to the Q model, the data is able to constrain two additional parameters, in agreement with~\cite{Zhao:2017cud,Wang:2018fng}.
For more complicated models, and in particular for simple phenomenological MG parametrizations~\cite{Abbott:2018xao}, we show that the data can constrain a significant number of extra degrees of freedom, that range from $10$ to $24$, depending on the model.

The determination of the number of effective parameters is expected to be accurate to a fraction of a parameter, as discussed in~\cite{Raveri:2018wln} and we performed all tests described there to ensure this.
Notice that the sum of the marginalized number of effective parameters is not guaranteed to be equal to the difference in total number with respect to $\Lambda$CDM because some of them might be correlated. As we will show in the next sections this effect is rather small and is mostly influenced by a small number of partially constrained parameter space directions.

We can also see from Tab.~\ref{Table:ReconstructionNeff} that $\alpha_K$ at late times is never constrained by the data.
As shown in~\cite{Bloomfield:2012ff} the relevant EFT operator drops out on small scales and is probed by horizon scale perturbations.
Since we are limiting the reconstruction to late times these horizon modes are very few in the data (i.e. CMB ISW modes) and have larger sensitivity to changes in the background than changes at the perturbation level.
For this reason we will not discuss further the role of $\alpha_K$ in the reconstruction.

% Evidence calculation
%
\begin{table}[!ht]
\setlength{\tabcolsep}{11pt}
\centering
\begin{tabular}{@{}llllll@{}}
\toprule
model & $\Delta \log_{10}\mathcal{E}$ &  $\Delta \log_{10}\mathcal{E}_{g}$ & $\Delta \log_{10}\mathcal{O}$\\
\toprule
Q   & $5.1$ & $4.8$ & $4.9$ \\
%\colrule
%K   & $5.2$ & $4.7$ & $4.9$ \\
\colrule
KGB & $10.9$ & $10.3$ & $10.2$ \\
\colrule
GBD & $18.5$ & $19.1$ & $18.4$ \\
\colrule
SH & $17.6$ & $17.5$ & $19.5$ \\
\colrule
FH & $25.9$ & $25.3$ & $27.0$ \\
\botrule
\end{tabular}
\caption{ \label{Table:ReconstructionEvidenceRatio}
The evidence ratio test for the EFT reconstruction against the $\Lambda$CDM model.
Note that a positive value of $\Delta \log_{10}\mathcal{E}$ denotes preference for the $\Lambda$CDM model. 
The first column shows its value computed from the MCMC samples while the second column shows its Gaussian approximation. The third column quantifies the Occam's razor penalty of the evidence comparison due to the increased prior volume of the reconstruction, Eq.~(\ref{Eq:OccamsRazor}).
Model acronyms are defined in Sec.~\ref{Sec:EFT}.
}
\end{table}

We then quantify the performances of the different models in fitting the data better than $\Lambda$CDM.
We start by considering the Bayesian evidence ratio test, as shown in Tab.~\ref{Table:ReconstructionEvidenceRatio}.
As we can see from the first two columns all reconstructed models are heavily penalized in the evidence comparison, showing results that are strongly disfavored on Jeffreys' scale.
While prior dominated parameter space directions drop out of the evidence calculation, we have seen in Tab.~\ref{Table:ReconstructionNeff} that the EFT reconstruction adds many parameters to $\Lambda$CDM that the data is sensitive to. 
That the evidence ratio is large shows that the new models are not fitting the data better than $\Lambda$CDM by a large enough factor to overcome the Occam's razor penalty of the constrained parameters they add. 
We can clearly gauge this penalty by comparing the first and third columns of Tab.~\ref{Table:ReconstructionEvidenceRatio}.
The latter column quantifies how much of the evidence ratio test is contributed by the Occam's razor factor which is dominating the result.

In Tab.~\ref{Table:ReconstructionEvidenceRatio} there is also remarkable agreement between the calculation of the evidence ratio from the MCMC samples and the results obtained with the Gaussian approximation.
The evidence calculation is known to be challenging in practice and the results obtained from the MCMC samples are robust because all the reconstruction chains have a large number of samples.
These results then show that the Gaussian approximation works well in describing both the prior and the posterior of the reconstruction for all the models.
Other quantities that we compute, based on the Gaussian approximation, should then be accurate as well.

% Best fitting model:
%
\begin{table}[!ht]
\setlength{\tabcolsep}{11pt}
\centering
\begin{tabular}{@{}llllll@{}}
\toprule
data set   & Q      & KGB    & GBD    & SH     & FH     \\
\toprule
CMB        & $-0.3$ & $+2.8$ & $-1.7$ & $-1.0$ & $+0.9$ \\
\colrule
BAO        & $+0.1$ & $-0.7$ & $+0.7$ & $+0.9$ & $+2.3$ \\
\colrule
SN         & $-0.1$ & $-0.7$ & $+0.5$ & $+1.6$ & $+2.9$ \\
\colrule
H0         & $-1.1$ & $+2.5$ & $+3.0$ & $+8.3$ & $+8.3$ \\
\colrule
WL         & $+3.4$ & $-3.0$ & $-1.2$ & $+3.2$ & $-1.5$ \\
\botrule
total      & $2.0$  & $0.9$  & $1.3$ & $13.0$ & $12.9$ \\
\colrule
SNR        & $1.4\sigma$ & $1\sigma$ & $1.1\sigma$ & $3.6\sigma$ & $3.6\sigma$ \\
\botrule
\end{tabular}
\caption{ \label{Table:ReconstructionMaxLike}
The data breakdown of the improvement in the best fit data $\chi^2$ value. Notice that a positive value denotes that the EFT reconstruction is performing better than the $\Lambda$CDM model.
Model acronyms are defined in Sec.~\ref{Sec:EFT}.
}
\end{table}

Since the evidence ratio comparison is usually unfair to the reconstruction and not indicative of the overall statistical significance that can be extracted from the data by a real model, as discussed in Sec.~\ref{Sec:ReconstructionMethod}, we now quantify whether the reconstructed models are providing a better fit to the data by comparing their values of the data likelihoods.
In Tab.~\ref{Table:ReconstructionMaxLike} we show the best fit data breakdown for each of the models.

Both FH and SH have similar performances and both are strongly preferred over $\Lambda$CDM cosmology.
The best fit solution slightly differs for some data sets. This is possibly due to the fact that the best fit is likely to lay on a degenerate direction in parameter space.
The common feature is that both models improve the fit to local measurements of the Hubble constant by a significant amount. 
They also improve the fit to BAO and SN measurements, in different amounts because of compromises with other data sets like the CMB and WL for which they give slightly different results.
We also note that changes the data likelihood are much smaller than the number of data points that each of the data sets contain and therefore that both models do not over-fit the data.

Even though both models show significant deviations from $\Lambda$CDM they do not entirely solve present tensions completely. In the previous section we roughly estimated that this would result in a $\chi^2$ improvement of about $22$ and both models achieve approximately half of that. 
In particular, regarding the tension in the determination of $H_0$, both models achieve about two third of the overall possible statistical significance.
Regarding the CMB lensing tension and the WL tension no models seem to provide a resolution.

While FH and SH are favored over $\Lambda$CDM in a best-fit sense, the more restricted Q, KGB and GBD models have similarly bad performances. 
All three models result in negligible preference with respect to the $\Lambda$CDM model.
In Sec.~\ref{Sec:Quintessence},~\ref{Sec:KGB} and~\ref{Sec:GBD} we comment on the physical reasons why this happens.

While we analyze the cosmology of the best fit models in the following sections, we comment here on the general trend of the changes to the parameters that are relevant for the discrepancy in the determination of the Hubble constant.

\begin{figure}[tbp]
\centering
\includegraphics[width=\columnwidth]{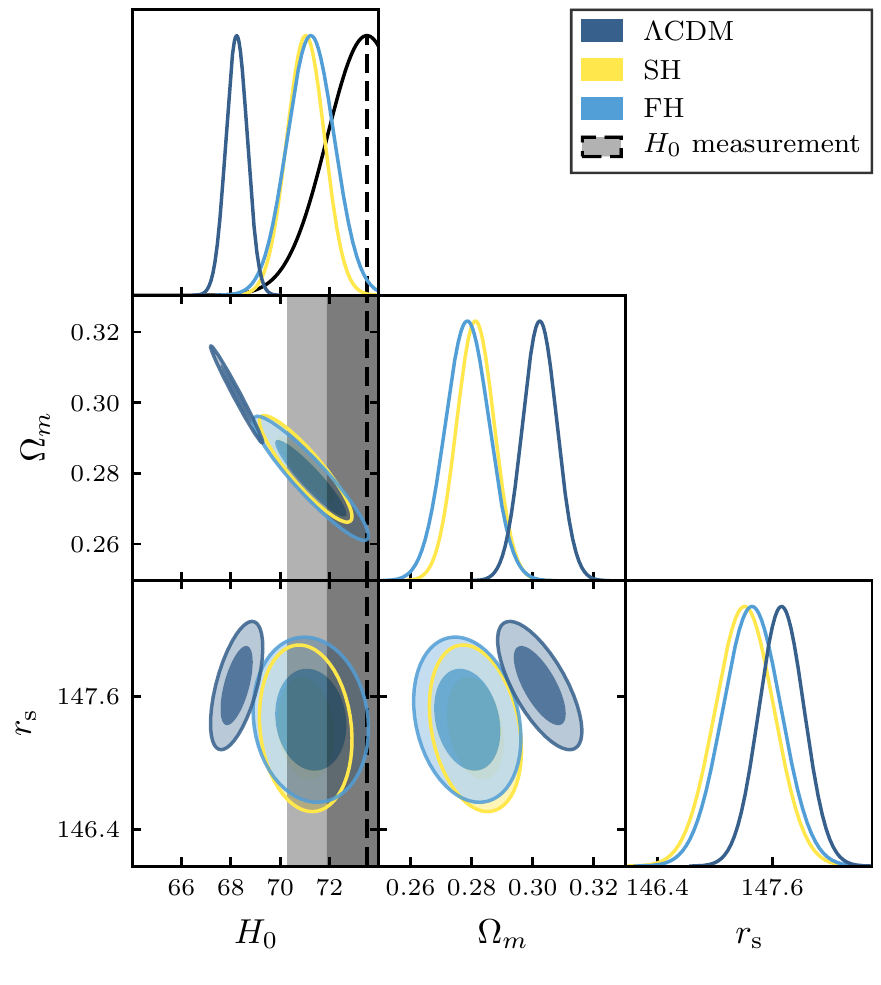}
\caption{ \label{fig:Triangle1} 
The joint marginalized posterior distribution of the Hubble constant $H_0$, matter density $\Omega_m$ and the scale of the sound horizon at the time of radiation drag $r_s$.
Different colors correspond to different models, as shown in legend. The darker and lighter shades correspond to the $68\%$ C.L. and $95\%$ C.L. regions respectively.
}
\end{figure} 

In Fig.~\ref{fig:Triangle1} we show the joint marginalized posterior distribution of the Hubble constant $H_0$, matter density $\Omega_m$ and the scale of the sound horizon at radiation drag $r_s$.
The inferred value of the sound horizon is only slightly changed with respect to the $\Lambda$CDM model and reflects small changes in other CMB parameters.
Our results can be used to gauge how much its determination can be changed by greatly relaxing assumptions on late time physics.
On the other hand the correlation between the sound horizon and the value of the Hubble constant is significantly weakened and the Hubble constant is allowed to shift to higher values because the inference of the low redshift BAO standard rulers, calibrated with SN, is changed by the significant shift in the inferred value of $\Omega_m$.

\begin{figure}[tbp]
\centering
\includegraphics[width=\columnwidth]{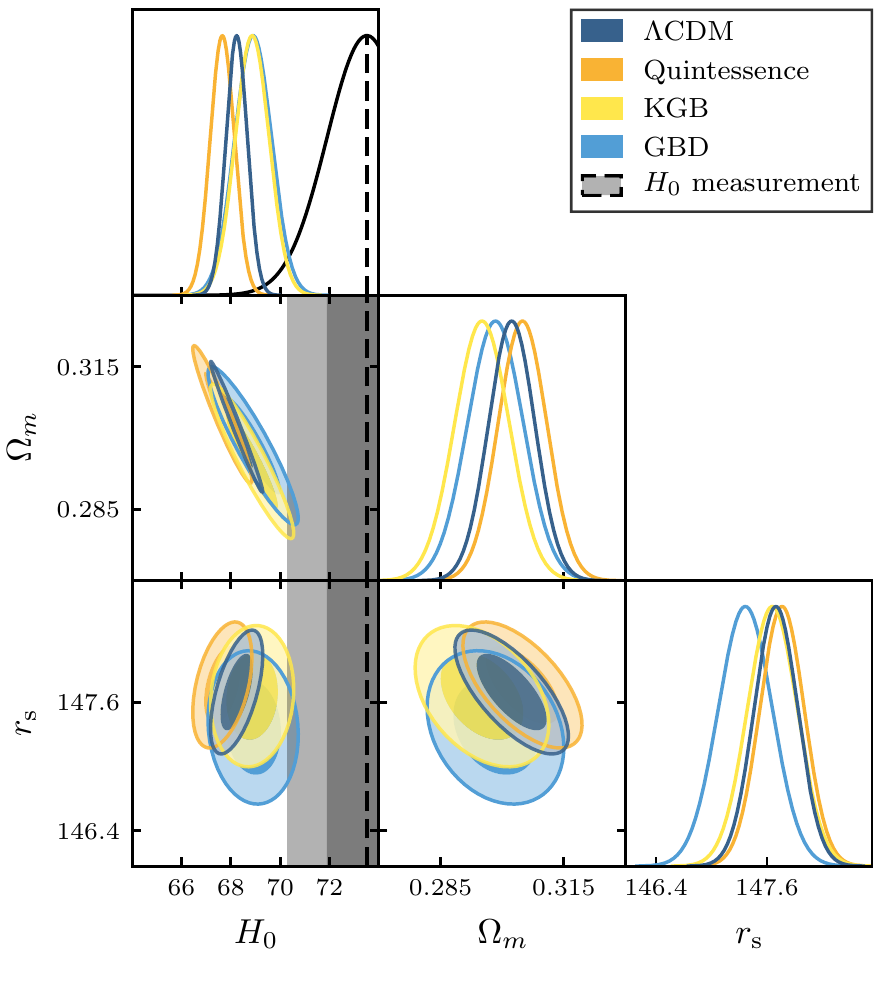}
\caption{ \label{fig:Triangle2} 
The joint marginalized posterior distribution of the Hubble constant $H_0$, matter density $\Omega_m$ and the scale of the sound horizon at radiation drag $r_s$.
Different colors correspond to different models, as shown in legend. The darker and lighter shades correspond to the $68\%$ C.L. and $95\%$ C.L. regions respectively.
}
\end{figure} 

In Fig.~\ref{fig:Triangle1} we show the posterior distribution for the models that do not provide a significant better fit to the $\Lambda$CDM model.
Quintessence models cannot raise the inferred value of the Hubble constant with respect to the $\Lambda$CDM model, as was also shown in~\cite{Raveri:2018ddi}.
Both KGB and GBD models can improve on the Quintessence results but not to a point that becomes largely statistically significant.

For further details on the reconstruction of the specific models as well as comments on the resulting cosmology we refer the reader to the following sub-sections.

%%%%%%%%%%%%%%%%%%%%%%%%%%%%%%%%%%%%%%%%%%%%%%%%%%%%%%%%%%%%%%%%%%%%%%%%
\subsection{Quintessence} \label{Sec:Quintessence}
%%%%%%%%%%%%%%%%%%%%%%%%%%%%%%%%%%%%%%%%%%%%%%%%%%%%%%%%%%%%%%%%%%%%%%%%
%
\begin{figure*}[!htbp]
\centering
\includegraphics[width=\textwidth]{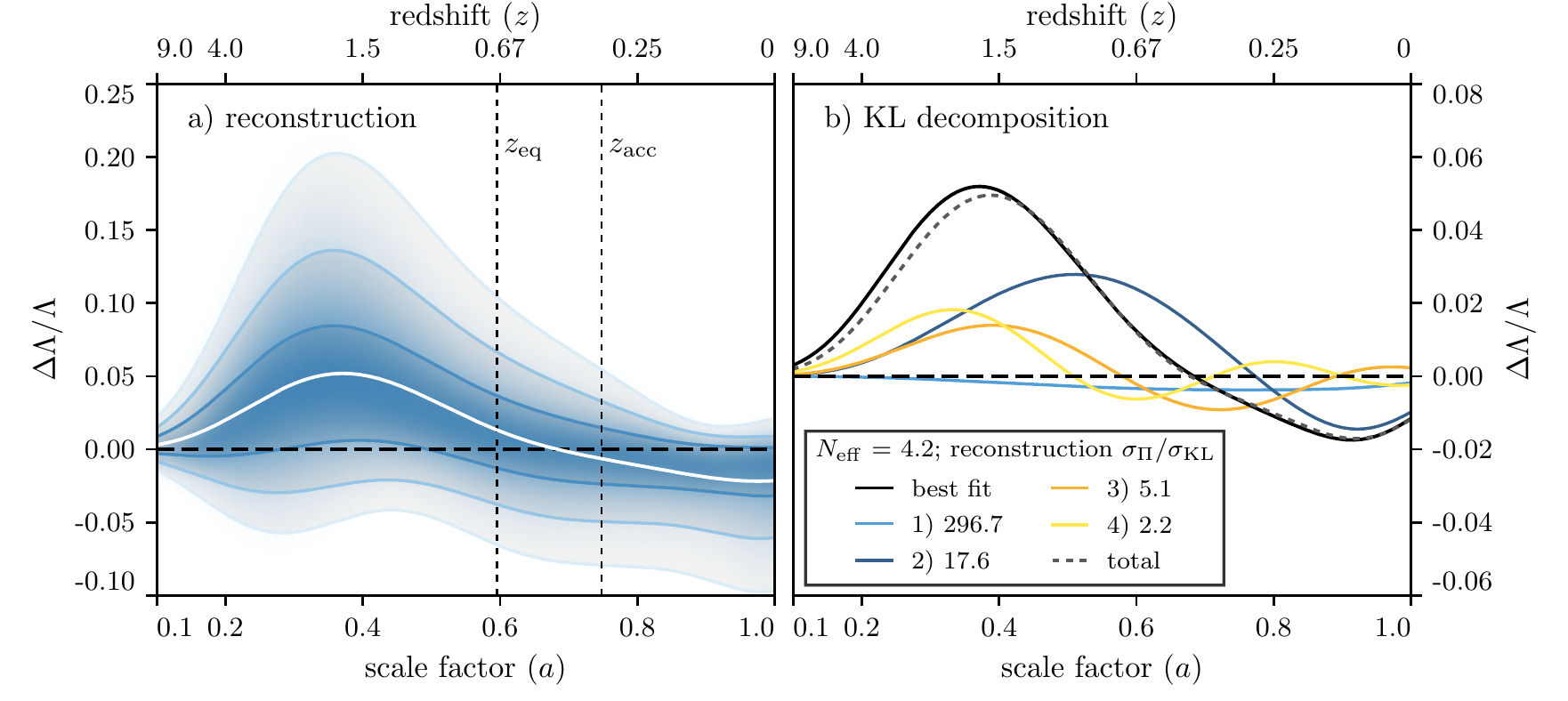}
\caption{ \label{fig:LambdaReconstruction} 
{\bf Reconstruction of Quintessence models.}
{\it Panel (a)} the marginalized posterior distribution of the EFT function $\Delta \Lambda/\Lambda$, describing all Quintessence models, as a function of scale factor and redshift.
The white line shows the mean of the distribution while the other contours represent the $68\%$, $95\%$ and $99.7\%$ C.L. regions respectively. The shade represents the posterior probability distribution. 
The two dashed lines show the redshift of equality between DM and DE, $z_{\rm eq}$, and the redshift of the beginning of cosmic acceleration $z_{\rm acc}$ in the best fitting $\Lambda$CDM model.
{\it Panel (b)} the KL decomposition of the best fitting Quintessence cosmological model. 
The continuous black line represent the best fit model.
Different lines correspond to different KL modes, as shown in legend.
The black dashed line shows the model obtained as the sum of the KL modes that are shown.
}
\end{figure*} 
\begin{figure}[!htb]
\centering
\includegraphics[width=\columnwidth]{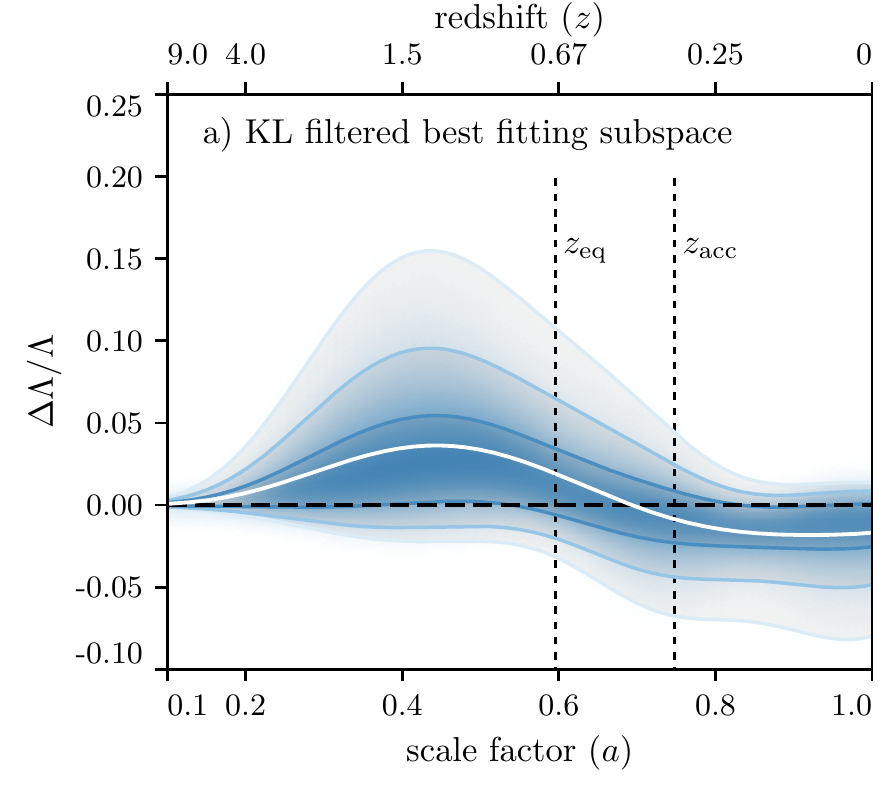}
\caption{ \label{fig:LambdaFilteredReconstruction} 
{\bf Reconstruction of Quintessence models.}
The marginalized distribution of the best fitting subspace for the reconstruction of Quintessence models.
All quantities follow the same conventions of Fig.~\ref{fig:LambdaReconstruction}. 
KL modes have been filtered by requiring that $\sigma_{\rm KL}<3 \sigma_{\Pi}$.
}
\end{figure} 
\begin{figure*}[!htbp]
\centering
\includegraphics[width=\textwidth]{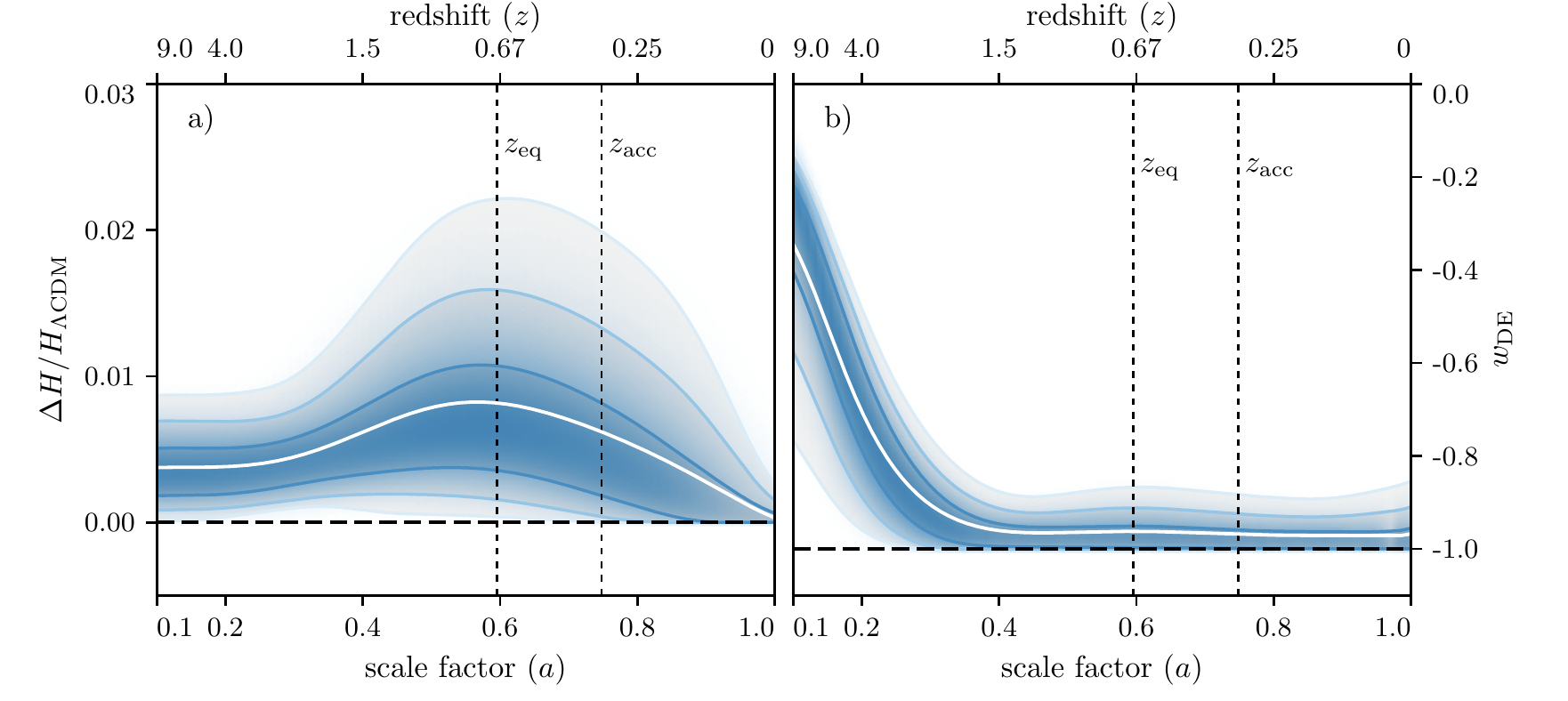}
\caption{ \label{fig:LambdaDerivedReconstruction} 
{\bf Reconstruction of Quintessence models.}
The marginalized distribution of variations in the expansion history, {\it Panel (a)}, and equation of state of DE, {\it Panel (b)}, in Quintessence models.
In both panels the white line represents the mean of the distribution while the other contours represent the $68\%$, $95\%$ and $99.7\%$ C.L. contours respectively.
The shade represents the probability distribution. The two dashed lines represent the redshift of equality between DM and DE and the redshift of the beginning of cosmic acceleration in the best fitting $\Lambda$CDM model.
}
\end{figure*} 
In this section we present the reconstruction of Quintessence (Q) models.

This is shown in Fig.~\ref{fig:LambdaReconstruction}\hyperref[fig:LambdaReconstruction]{a} in terms of the EFT function $\Delta\Lambda/\Lambda$. The limit of this model to the $\Lambda$CDM one is represented by $\Delta\Lambda/\Lambda=0$ at all times.
The reconstruction is consistent with the $\Lambda$CDM model at all times reflecting the fact that the model is not providing a significantly better fit to the data.
The overall constraints are at the $10\%$ level after the redshift of DM/DE equality and increase at higher redshift.

The main cosmological effect of this model is that of changing the rate of dilution of DE density, and correspondingly increasing the expansion rate, in the past. 
At the perturbation level the model is in the class of smooth DE models and changes the overall amplitude of perturbations in a scale independent way.
The increasingly lower constraining power in the past reflects the fact that the DE component is progressively less dominant and results in smaller effects on the expansion history of the model.
At about $a=0.3$ the constraints start shrinking reflecting the fact that the model has to be smoothly joined to its $\Lambda$CDM limit at $a=0.1$.

The constraints shown in Fig.~\ref{fig:LambdaReconstruction}\hyperref[fig:LambdaReconstruction]{a} are marginalized over a very wide parameter space that includes functional modes that are not well constrained by the data.
To single out the role of the data in constraining this model we study the KL decomposition of the reconstruction, discussed in Sec.~\ref{Sec:ReconstructionMethod}.
In Fig.~\ref{fig:LambdaReconstruction}\hyperref[fig:LambdaReconstruction]{b} we show the KL modes that are constrained by the data over the prior.
The number of modes singled out matches the number of effective parameters that the data is constraining $N_{\rm eff}=4.2$, as in Tab.~\ref{Table:ReconstructionNeff}.
The first two of these modes are strongly constrained by the data and correspond to the smoothest variations in time.
These two modes are the ones constrained in phenomenological approaches to these types of models, like the $(w_0,w_a)$ CPL parametrization.
The third mode is still being constrained and the prior is not informative below the five sigma level which means that the CPL parametrization is not lossless by at least one parameter. The last mode corresponds to faster variations and is the least constrained one.
The structure of the KL modes is oscillatory because the data constraints are integrals over the expansion history.
We also notice that the shape of the KL modes would significantly differ from that of the PC of the posterior covariance.

We then project the reconstruction over these best data constrained modes to single out the best fitting subspace that is shown in Fig.~\ref{fig:LambdaFilteredReconstruction}.
Notice that the constraints shown are, as discussed in Sec.~\ref{Sec:ReconstructionMethod}, slightly tighter than the reconstruction ones as they show the minimal space where the model would provide a good fit to the data.
As we can clearly see now the $68\%$ C.L. contour includes the $\Lambda$CDM limit of the model, reflecting the fact that the model is preferred by the data at the $1\sigma$ level.

The reconstruction fully determines the properties of the model that is being studied and we can use it to project the constraints over model properties as we show in Fig.~\ref{fig:LambdaDerivedReconstruction}.

In Panel (\hyperref[fig:LambdaDerivedReconstruction]{a}) we see variations with respect to the corresponding $\Lambda$CDM expansion history.
Notice that these are not variations with respect to the best fit $\Lambda$CDM expansion history but are computed for the same choice of parameters, model by model.
Variations is expansion history are always positive. Since the equation of state of the DE component cannot cross $w=-1$, as we can see from Panel (\hyperref[fig:LambdaDerivedReconstruction]{b}), because of ghost instabilities, then, at fixed energy density today, the energy density of DE in the past has to be larger than the corresponding $\Lambda$CDM model and accordingly the expansion rate has to be larger.
This is the reason why the model, even in full generality, cannot explain the discrepancy between the CMB and Hubble constant measurements and is penalized accordingly.
The maximum deviation from the $\Lambda$CDM expansion history is at most $5\%$ (at $99.7\%$ C.L.) at the redshift of DM/DE equality. At times earlier than that it decays because the field becomes subdominant.

In Panel (\hyperref[fig:LambdaDerivedReconstruction]{b}) we can see the constraints on the equation of state. When the model smoothly matches the $\Lambda$CDM limit the equation of state tends to zero.
This is a property of the EFT basis that we are considering, as discussed in~\cite{Raveri:2017qvt}. When the model approaches the $\Lambda$CDM limit the DE field becomes a tracker field that follows the equation of state of DM.
Notice that by the time the field is smoothly going to the tracker solution its energy density is largely subdominant with respect to the DM one and has no observational consequence.
After equality the equation of state of DE is constrained to be below $w<-0.85$ (at $99.7\%$ C.L.) at all times, within Quintessence models.

%%%%%%%%%%%%%%%%%%%%%%%%%%%%%%%%%%%%%%%%%%%%%%%%%%%%%%%%%%%%%%%%%%%%%%%%
\subsection{Kinetic Gravity Braiding} \label{Sec:KGB}
%%%%%%%%%%%%%%%%%%%%%%%%%%%%%%%%%%%%%%%%%%%%%%%%%%%%%%%%%%%%%%%%%%%%%%%%
%
\begin{figure*}[!htbp]
\centering
\includegraphics[width=\textwidth]{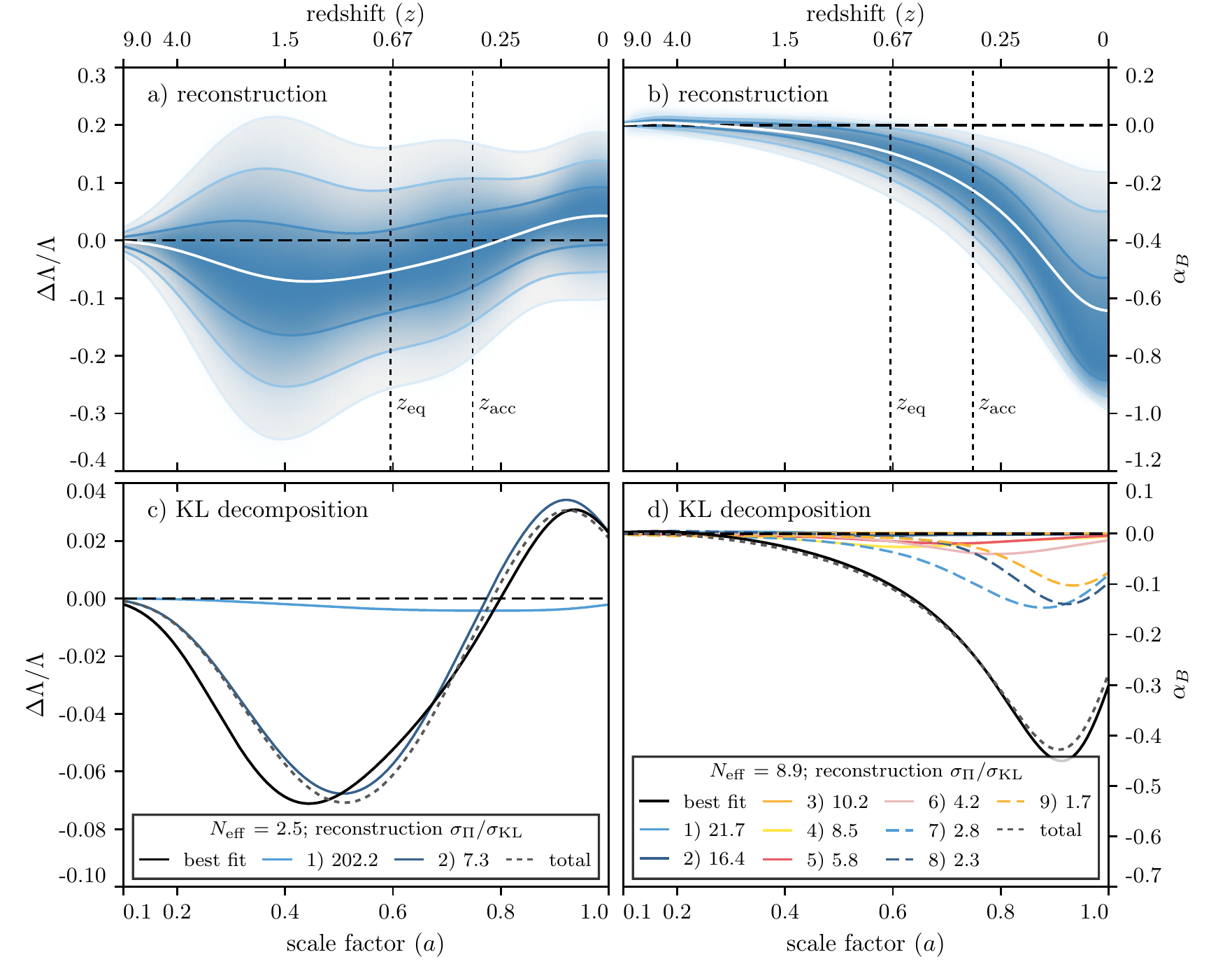}
\caption{ \label{fig:KGBReconstruction} 
{\bf Reconstruction of Kinetic Gravity Braiding models.}
{\it Panels (a,b)} the marginalized posterior distribution of the EFT function $\Delta \Lambda/\Lambda$ and $\alpha_B$, describing all Kinetic Gravity Braiding (KGB) models, as a function of scale factor and redshift.
The white line shows the mean of the distribution while the other contours represent the $68\%$, $95\%$ and $99.7\%$ C.L. regions respectively. The shade represents the posterior probability distribution. 
The two dashed lines show the redshift of equality between DM and DE, $z_{\rm eq}$, and the redshift of the beginning of cosmic acceleration $z_{\rm acc}$ in the best fitting $\Lambda$CDM model.
{\it Panels (c,d)} the KL decomposition of the best fitting KGB cosmological model. 
The continuous black line represent the best fit model.
Different lines correspond to different KL modes, as shown in legend.
The black dashed line shows the model obtained as the sum of the KL modes that are shown.
}
\end{figure*} 
\begin{figure*}[p!htb]
\centering
\includegraphics[width=\textwidth]{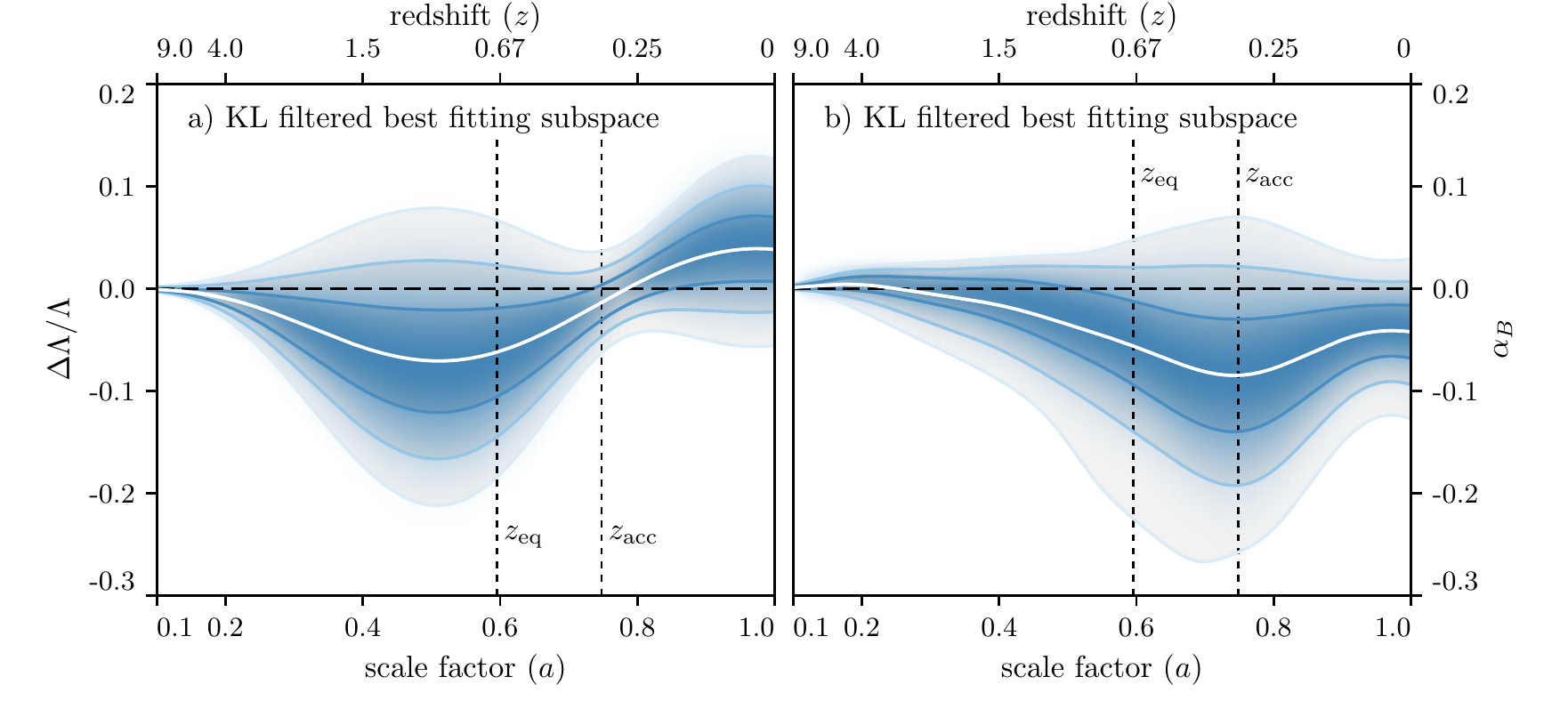}
\caption{ \label{fig:KGBFilteredReconstruction} 
{\bf Reconstruction of Kinetic Gravity Braiding models.}
The marginalized distribution of the best fitting subspace for the reconstruction of Kinetic Gravity Braiding models.
The white line shows the mean of the distribution while the other contours represent the $68\%$, $95\%$ and $99.7\%$ C.L. regions respectively. The shade represents the posterior probability distribution. 
The two dashed lines show the redshift of equality between DM and DE, $z_{\rm eq}$, and the redshift of the beginning of cosmic acceleration $z_{\rm acc}$ in the best fitting $\Lambda$CDM model.
KL modes have been filtered by requiring that $\sigma_{\rm KL}<3 \sigma_{\Pi}$.
}
\vspace{1cm}
\includegraphics[width=\textwidth]{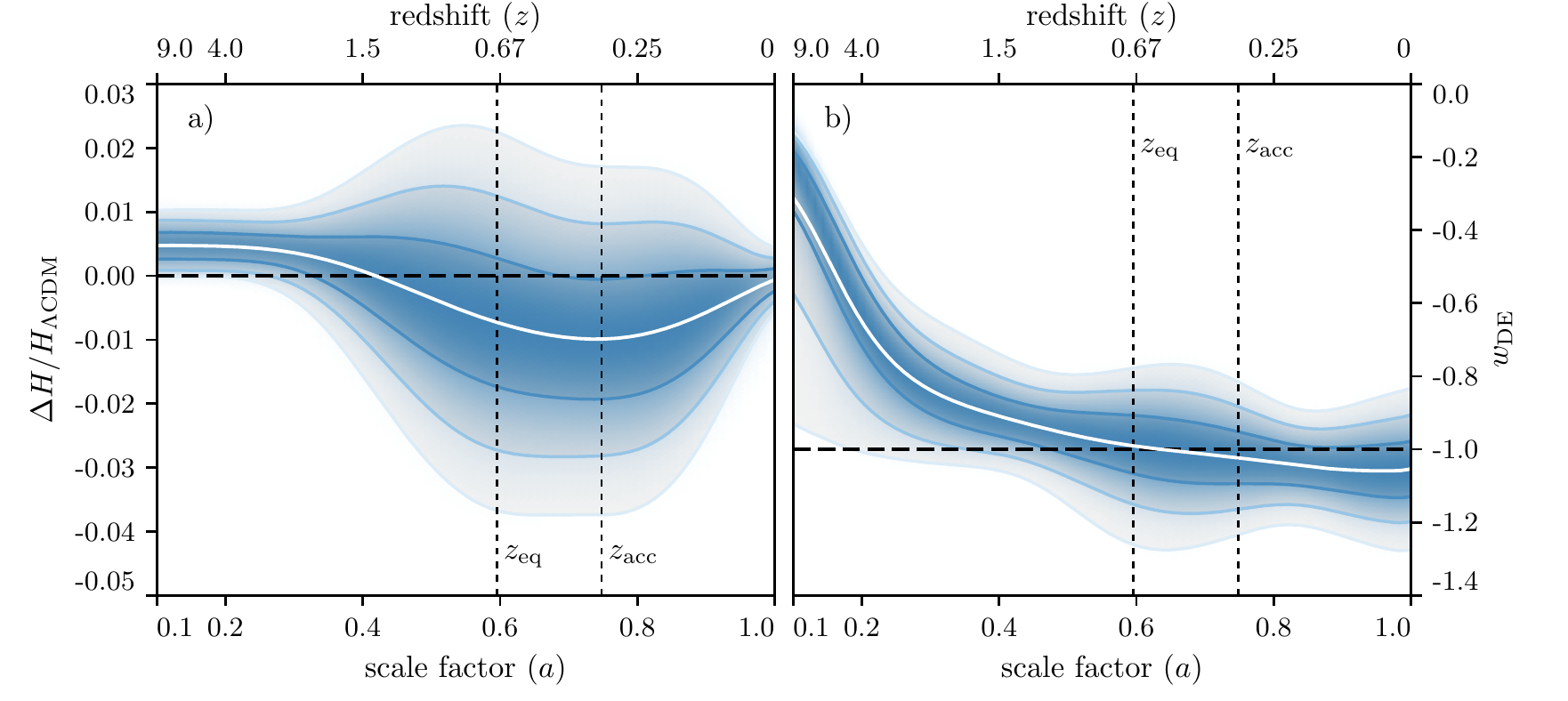}
\caption{ \label{fig:KGBDerivedReconstruction} 
{\bf Reconstruction of Kinetic Gravity Braiding models.}
The marginalized distribution of variations in the expansion history, {\it Panel (a)}, and equation of state of DE, {\it Panel (b)}, in Kinetic Gravity Braiding models.
In both panels the white line represents the mean of the distribution while the other contours represent the $68\%$, $95\%$ and $99.7\%$ C.L. contours respectively.
The shade represents the probability distribution. The two dashed lines represent the redshift of equality between DM and DE and the redshift of the beginning of cosmic acceleration in the best fitting $\Lambda$CDM model.
}
\end{figure*} 

In this section we present the reconstruction of Kinetic Gravity Braiding (KGB) models.
This is shown in Fig.~\ref{fig:KGBReconstruction}\hyperref[fig:KGBReconstruction]{a,b} in terms of the EFT function $\Delta\Lambda/\Lambda$ and $\alpha_B$. The limit of this model to the $\Lambda$CDM one is represented by $\Delta\Lambda/\Lambda=0$ and $\alpha_B=0$ at all times.

The reconstruction of $\Delta\Lambda/\Lambda$ is consistent with the $\Lambda$CDM model at all times while the reconstructed $\alpha_B$ appears to be highly discrepant.
As we will show later this is not truly representative of the fact that the model is preferred to the $\Lambda$CDM model but is rather a reflection of the large prior volume of the reconstruction.
The KGB model is, in fact, not providing a better fit to the data, as shown in Sec.~\ref{Sec:GeneralResults}.
The overall constraints are at the $20\%$ level after the redshift of DM/DE for $\Delta\Lambda/\Lambda$ and can be $100\%$ for $\alpha_B$.

With respect to the $\Lambda$CDM model, the main cosmological effect of KGB, at the background level, is that of changing the expansion rate.
At the perturbation level, differently from Q models, inhomogeneities in the DE field can grow in a scale dependent way since $\alpha_B$ changes their effective sound horizon.
This means that, for suitably chosen $\alpha_B$, DE perturbations can cluster on sub-sound horizon scales.
Notice that $\alpha_B$ does not enter in the definition of the cosmological background but only affects the behavior of perturbations.
Still the increasingly lower constraining power in the past for $\Delta\Lambda/\Lambda$ reflects the fact that the DE component is progressively less dominant and results in smaller overall effects.
On the other hand $\alpha_B$ seems to be more constrained at early times but this is just because of the presence of stability priors.
We verified, in fact, that this behavior is already present in the prior only chains and set by the fact that the model should not display gradient instabilities that would correspond to a negative effective sound speed.

As in the previous model, at about $a=0.3$ the constraints on 
$\Delta\Lambda/\Lambda$ start shrinking reflecting the fact that the model has to be smoothly joined to its $\Lambda$CDM limit at $a=0.1$.

Since the model displays a marked influence of the prior volume, that is marginalized over for the weakly constrained modes we turn to the KL modes analysis, shown in Fig.~\ref{fig:KGBReconstruction}\hyperref[fig:KGBReconstruction]{c,d}. 
As we can see, for both EFT functions, the number of KL modes matches the number of effective parameters that the data is constraining, as in Tab.~\ref{Table:ReconstructionNeff}.
The total number of constrained parameters is much larger than in the previous model, reflecting the fact that perturbations are now contributing their constraining power. 

The number of KL modes that are constrained for the $\Delta\Lambda/\Lambda$ function is decreased with respect to the Q case to two well constrained modes.
This means that, in the Q model, some constraining power was contributed by perturbation observables, mainly the amplitude of the CMB lensing and WL power spectra.
On the other hand $\alpha_B$ has plenty of constrained modes.
At least the first five are highly constrained clearly showcasing the power of the late times probes of fluctuations.
The sum of the constrained KL modes of $\alpha_B$ matches the best fit behavior but the largest modes contributing to the total are the three ones that are worse constrained and therefore likely to be polluted by the prior.

We then project the reconstruction over the best constrained modes, for which $\sigma_{\rm KL}<3\sigma_{\Pi}$, to single out the data features in the best fitting subspace, shown in Fig.~\ref{fig:KGBFilteredReconstruction}.
This procedure truly unveils the data behavior that is now noticeably different from the posterior one.
As explained in the previous section the constraints become tighter, reflecting the fact that this figure is representing the minimal subspace, with its uncertainty, where the model would provide the better fit to the data.
As we can see now the reconstruction is largely compatible with the $\Lambda$CDM model, as the KGB model is preferred by the data at about the $1\sigma$ level.
This procedure also unveils some interesting features.
As we can see the zero crossing of $\Delta\Lambda/\Lambda$ happens at about the redshift of acceleration, where also the uncertainty in $\alpha_B$ is peaked. No noticeable event seems related to the redshift of equality, pointing toward the fact that, in the KGB class of models compatible with observations, the redshift of acceleration is playing a key role. 

As in the previous section we now project the constraints over model properties that we show in Fig.~\ref{fig:KGBDerivedReconstruction}.
In Panel (\hyperref[fig:KGBDerivedReconstruction]{a}) we see variations with respect to the $\Lambda$CDM expansion history while in Panel (\hyperref[fig:KGBDerivedReconstruction]{b}) we show the DE equation of state.

As we can see the equation of state is now allowed to cross $w=-1$ because ghost instabilities in the Q model are now stabilized by the presence of $\alpha_B$.
Variations in the expansion history can now become smaller than zero, corresponding to models that have an energy density that dilutes in the past.
However, when the model needs to join its $\Lambda$CDM limit, variations in the expansion history are forced to become positive again. For this reason the model is overall penalized by the tension between the CMB and Hubble constant measurements.
This clearly points toward the fact that, in this model, the Hubble constant discrepancy may only be solved by considering models that have large deviations from $\Lambda$CDM around recombination.
The maximum deviation from the $\Lambda$CDM expansion history is at most $8\%$ (at $99.7\%$ C.L.) at the redshift of acceleration.
In addition the constraints on the equation of state, as a function of time, become now much looser and are now $-1.3<w_{\rm DE}<0.8$ between equality and today.

%%%%%%%%%%%%%%%%%%%%%%%%%%%%%%%%%%%%%%%%%%%%%%%%%%%%%%%%%%%%%%%%%%%%%%%%
\subsection{Generalized Brans-Dicke} \label{Sec:GBD}
%%%%%%%%%%%%%%%%%%%%%%%%%%%%%%%%%%%%%%%%%%%%%%%%%%%%%%%%%%%%%%%%%%%%%%%%
%
\begin{figure*}[!htbp]
\centering
\includegraphics[width=0.93\textwidth]{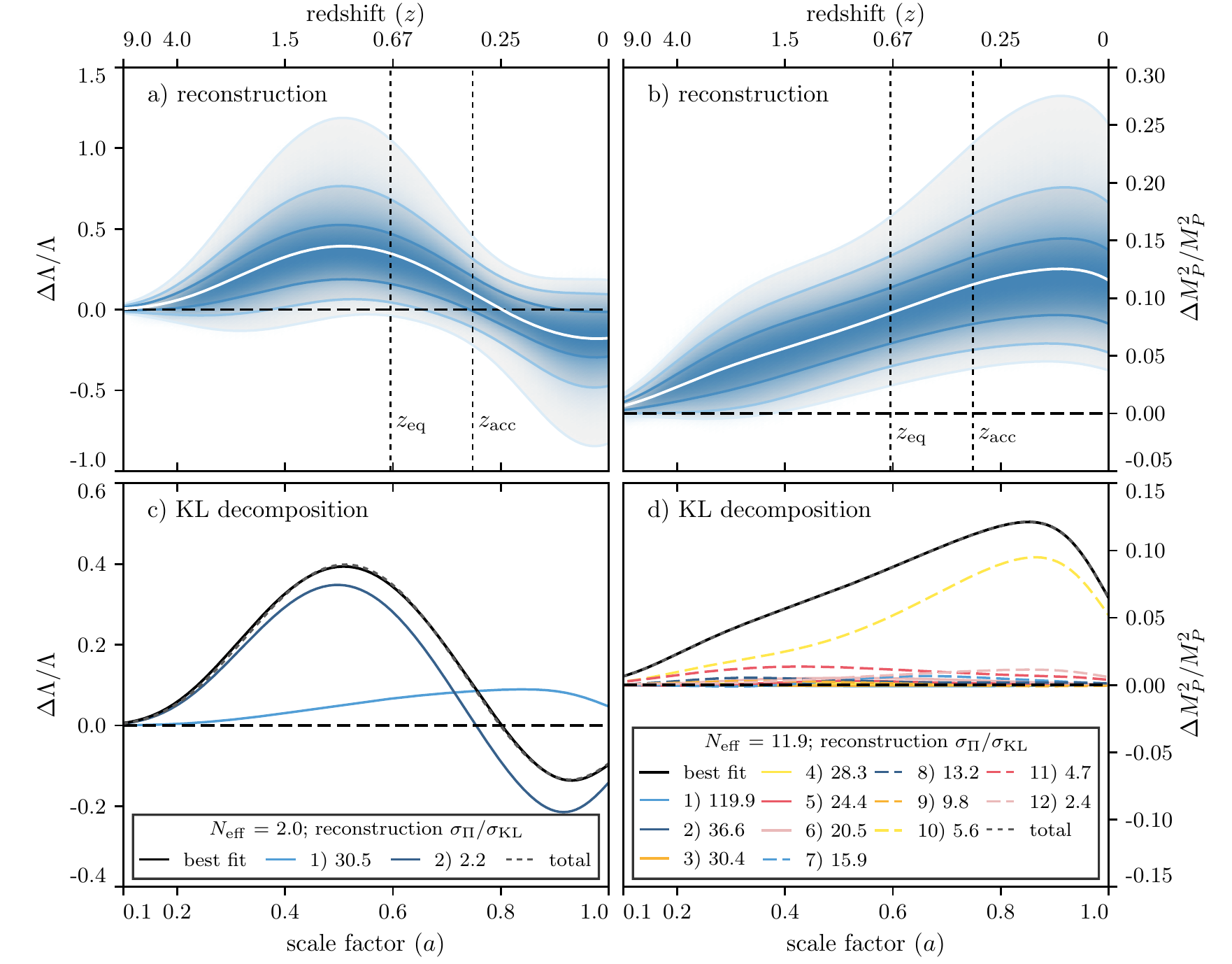}
\caption{ \label{fig:GBDReconstruction} 
{\bf Reconstruction of Generalized Brans-Dicke models.}
{\it Panels (a,b)} the marginalized posterior distribution of the EFT function $\Delta \Lambda/\Lambda$ and $\Delta M_P^2/M_P^2$, describing all Generalized Brans-Dicke (GBD) models, as a function of scale factor and redshift.
The white line shows the mean of the distribution while the other contours represent the $68\%$, $95\%$ and $99.7\%$ C.L. regions respectively. The shade represents the posterior probability distribution. 
The two dashed lines show the redshift of equality between DM and DE, $z_{\rm eq}$, and the redshift of the beginning of cosmic acceleration $z_{\rm acc}$ in the best fitting $\Lambda$CDM model.
{\it Panels (c,d)} the KL decomposition of the best fitting GBD cosmological model. 
The continuous black line represent the best fit model.
Different lines correspond to different KL modes, as shown in legend.
The black dashed line shows the model obtained as the sum of the KL modes that are shown.
}
\end{figure*} 
\begin{figure*}[!htbp]
\centering
\includegraphics[width=\textwidth]{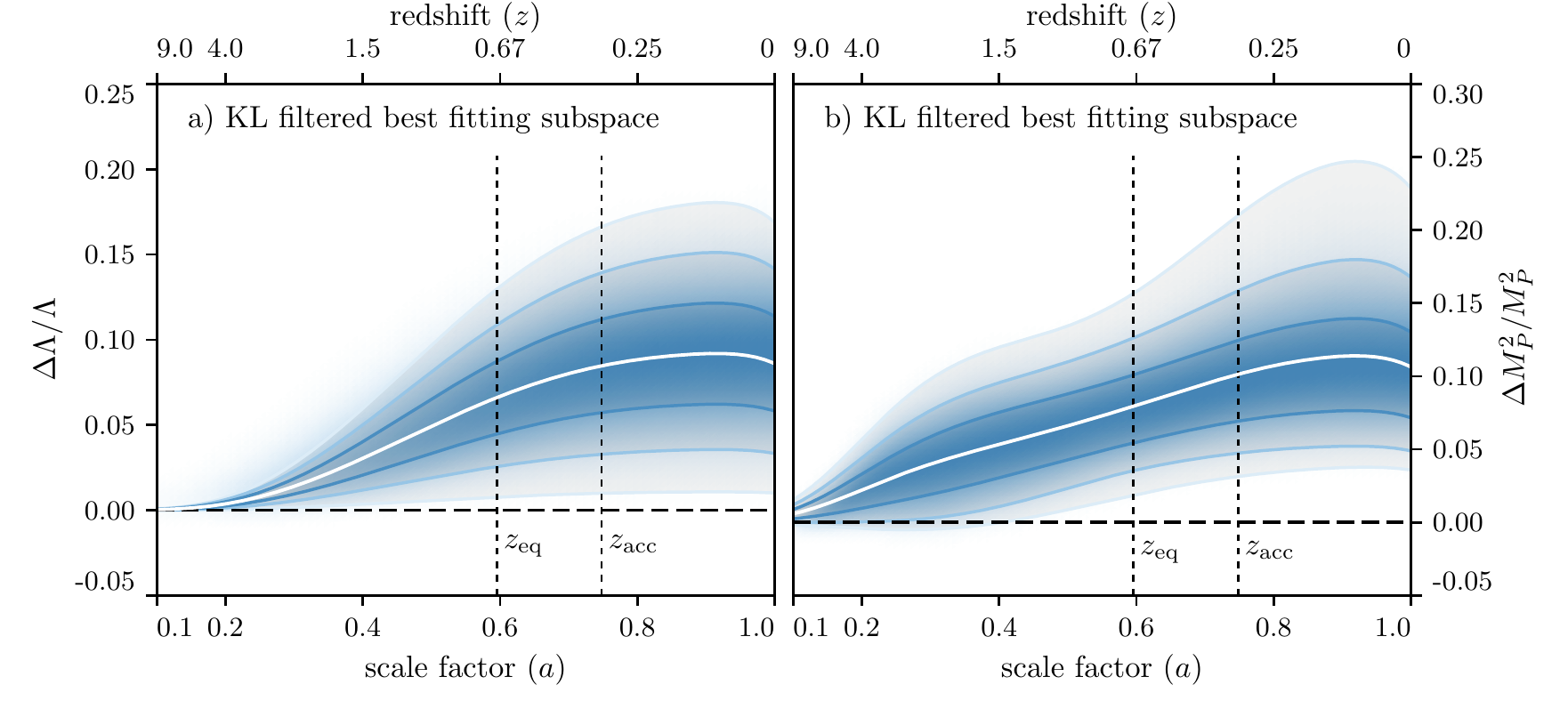}
\caption{ \label{fig:GBDFilteredReconstruction} 
{\bf Reconstruction of Generalized Brans-Dicke models.}
The marginalized distribution of the best fitting subspace for the reconstruction of Generalized Brans-Dicke models.
The white line shows the mean of the distribution while the other contours represent the $68\%$, $95\%$ and $99.7\%$ C.L. regions respectively. The shade represents the posterior probability distribution. 
The two dashed lines show the redshift of equality between DM and DE, $z_{\rm eq}$, and the redshift of the beginning of cosmic acceleration $z_{\rm acc}$ in the best fitting $\Lambda$CDM model.
KL modes have been filtered by requiring that $\sigma_{\rm KL}<3 \sigma_{\Pi}$.
}
\vspace{1cm}
\includegraphics[width=\textwidth]{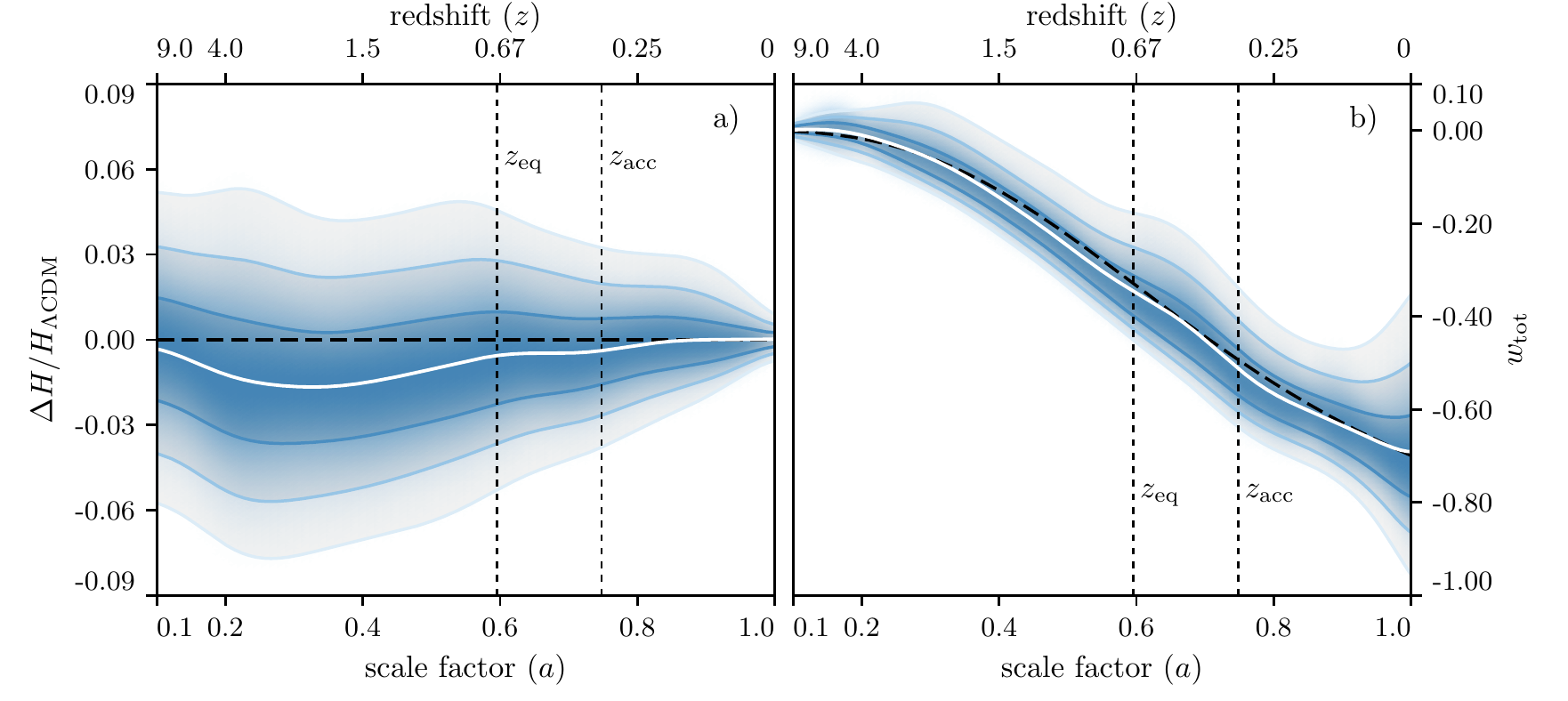}
\caption{ \label{fig:GBDDerivedReconstruction} 
{\bf Reconstruction of Generalized Brans-Dicke models.}
The marginalized distribution of variations in the expansion history, {\it Panel (a)}, and in the total effective equation of state, {\it Panel (b)}, in Generalized Brans-Dicke models.
In both panels the white line represents the mean of the distribution while the other contours represent the $68\%$, $95\%$ and $99.7\%$ C.L. contours respectively.
The shade represents the probability distribution. The two dashed lines represent the redshift of equality between DM and DE and the redshift of the beginning of cosmic acceleration in the best fitting $\Lambda$CDM model.
}
\end{figure*} 

In this section we present the reconstruction of Generalized Brans-Dicke (GBD) models.
This is shown in Fig.~\ref{fig:GBDReconstruction}\hyperref[fig:GBDReconstruction]{a,b} in terms of the EFT functions $\Delta\Lambda/\Lambda$ and $\Delta M_P^2/M_P^2$. The limit of this model to the $\Lambda$CDM one is represented by $\Delta\Lambda/\Lambda=0$ and $\Delta M_P^2/M_P^2=0$ at all times.

In contrast to the previous model the two EFT functions in GBD modify both the background and the behavior of perturbations.
At the background level $\Delta\Lambda/\Lambda$ enters as an additive term whose relevance increases as we move from DM domination to DE domination.
In contrast $\Delta M_P^2/M_P^2$ acts as a multiplicative term rescaling the overall expansion history at all times and in particular during matter domination too.
At the perturbation level $\Delta\Lambda/\Lambda$ has a limited effect while $\Delta M_P^2/M_P^2$ induces a scale dependent growth of perturbations.

We can now see that both the reconstruction of $\Delta\Lambda/\Lambda$ and $\Delta M_P^2/M_P^2$ appear to be deviating from the $\Lambda$CDM model. In particular deviations in $\Delta M_P^2/M_P^2$ seem to be strongly significant. 
As we have shown in Sec.~\ref{Sec:GeneralResults}, the GBD model does not provide a better fit to the data.
We verified that this trend is already present in the prior that show that $\Delta M_P^2/M_P^2$ is increasing with time. The prior then contains many models where this increase is large but very few where it is constant so the posterior is pushed toward higher values as a result of marginalization.

We compute the KL mode decomposition of both functions that is shown in Fig.~\ref{fig:GBDReconstruction}\hyperref[fig:GBDReconstruction]{c,d}.
The first feature to observe is that $\Delta\Lambda/\Lambda$ shows a larger degree of degeneracy with $\Delta M_P^2/M_P^2$ than with $\alpha_B$.
As in the KGB model only two modes are constrained by the data for $\Delta\Lambda/\Lambda$ but the constraints are much weaker, indicating a larger degree of degeneracy.
On the other hand $\Delta M_P^2/M_P^2$ displays a large number of constrained modes.
This is higher with respect to the KGB reconstruction because $\Delta M_P^2/M_P^2$ harvests constraining power from both the background and perturbations. 
At the background level, in particular, all the distance ladder measurements contribute their constraints, in particular during matter domination.
On the other hand modes that are constrained by perturbations get their constraining power from the fact that time dependencies get projected over scale dependencies once integrated over the window function of a given probe. These modes are then constrained by all redshift bins in WL and by their cross correlation and by the CMB lensing reconstruction.
As we can see, the KL mode that is contributing the most to the general trend in $\Delta M_P^2/M_P^2$ is still constrained to a significant amount.
When we project on the best fitting subspace, as shown in Fig.~\ref{fig:GBDFilteredReconstruction}, the overall behavior does not change significantly.
We can then interpret the volume effect that drives $\Delta M_P^2/M_P^2$ away from its GR limit as a non-trivial boundary against which the constraints are pushed.

We now project the reconstruction on derived cosmological quantities.
Since, in GBD, the conformal factor is non-standard we cannot define a meaningful equation of state of the effective DE fluid as it would become, for some models within the samples, ill-defined.
We indeed verified that many models would have discontinuities in the effective DE equation of state and this would not allow us to estimate the posterior probability distribution.
We then turn to considering the total effective equation of state.
In Fig.~\ref{fig:GBDDerivedReconstruction}\hyperref[fig:GBDDerivedReconstruction]{a}, we see variations with respect to the $\Lambda$CDM expansion history while in Panel (\hyperref[fig:GBDDerivedReconstruction]{b}) we show the total effective equation of state.
As we can see the fractional change in expansion history can now be negative at all times, which is a feature that KGB models could not achieve.
Correspondingly the total equation of state is allowed to be smaller than the $\Lambda$CDM one.
These changes are nonetheless insufficient to solve the discrepancy in the determination of the Hubble constant.
We see no particular trend, for this model, related to the redshift of equality or acceleration.

\begin{figure*}[!htbp]
\centering
\includegraphics[width=\textwidth]{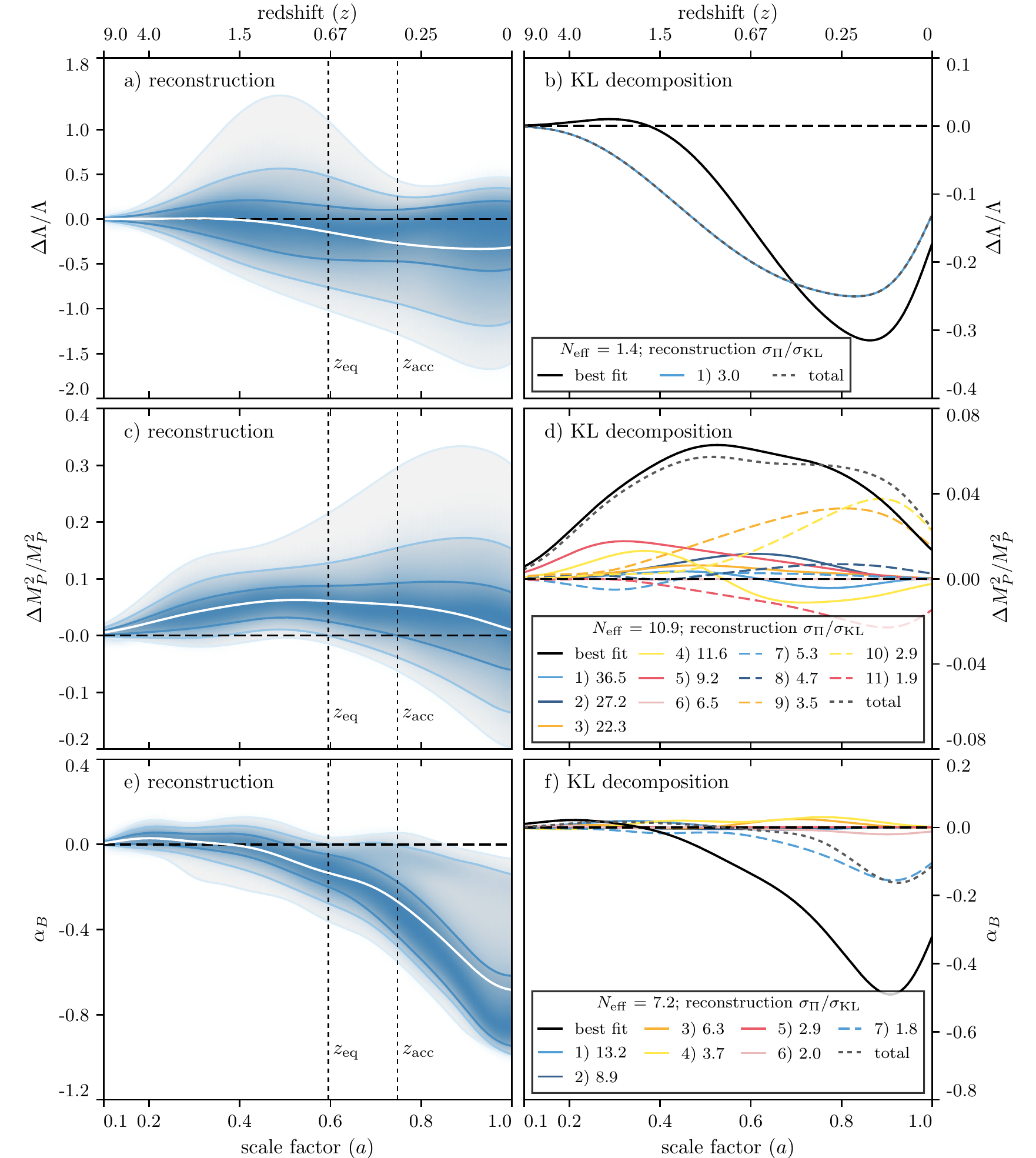}
\caption{ \label{fig:SHReconstruction} 
{\bf Reconstruction of Scalar Horndeski models.}
{\it Panels (a,c,e)} the marginalized posterior distribution of the EFT function $\Delta \Lambda/\Lambda$, $\Delta M_P^2/M_P^2$ and $\alpha_B$, describing all Scalar Horndeski (SH) models, as a function of scale factor and redshift.
The white line shows the mean of the distribution while the other contours represent the $68\%$, $95\%$ and $99.7\%$ C.L. regions respectively. The shade represents the posterior probability distribution. 
The two dashed lines show the redshift of equality between DM and DE, $z_{\rm eq}$, and the redshift of the beginning of cosmic acceleration $z_{\rm acc}$ in the best fitting $\Lambda$CDM model.
{\it Panels (b,d,f)} the KL decomposition of the best fitting SH cosmological model. 
The continuous black line represent the best fit model.
Different lines correspond to different KL modes, as shown in legend.
The black dashed line shows the model obtained as the sum of the KL modes that are shown.
}
\end{figure*} 

\clearpage

%%%%%%%%%%%%%%%%%%%%%%%%%%%%%%%%%%%%%%%%%%%%%%%%%%%%%%%%%%%%%%%%%%%%%%%%
\subsection{Scalar Horndeski} \label{Sec:SH}
%%%%%%%%%%%%%%%%%%%%%%%%%%%%%%%%%%%%%%%%%%%%%%%%%%%%%%%%%%%%%%%%%%%%%%%%

%
\begin{figure}[!htbp]
\centering
\includegraphics[width=\columnwidth]{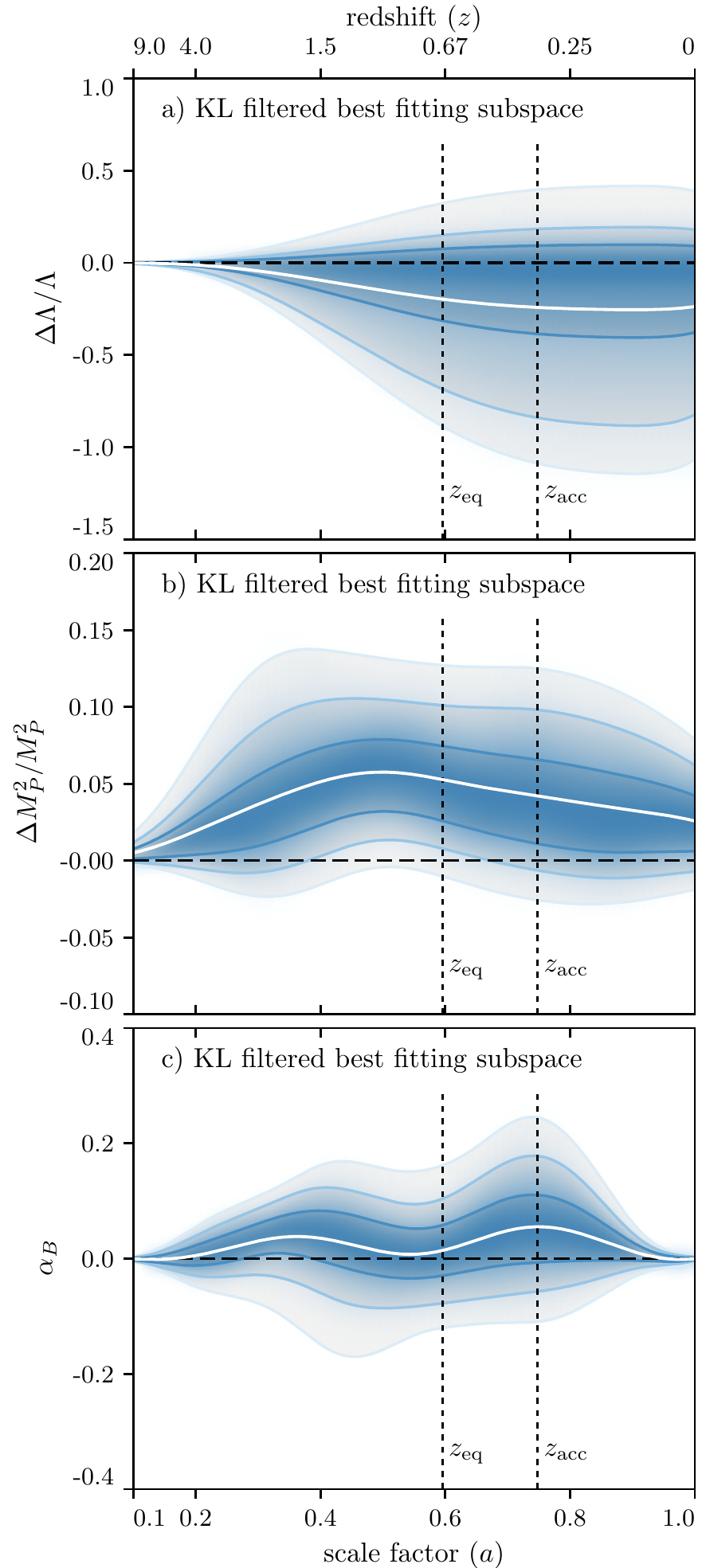}
\caption{ \label{fig:SHFilteredReconstruction} 
{\bf Reconstruction of Scalar Horndeski models.}
The marginalized distribution of the best fitting subspace for the reconstruction of Scalar Horndeski Braiding models.
The white line shows the mean of the distribution while the other contours represent the $68\%$, $95\%$ and $99.7\%$ C.L. regions respectively. The shade represents the posterior probability distribution. 
The two dashed lines show the redshift of equality between DM and DE, $z_{\rm eq}$, and the redshift of the beginning of cosmic acceleration $z_{\rm acc}$ in the best fitting $\Lambda$CDM model.
KL modes have been filtered by requiring that $\sigma_{\rm KL}<3 \sigma_{\Pi}$.
}
\end{figure} 
\begin{figure*}[!htbp]
\centering
\includegraphics[width=\textwidth]{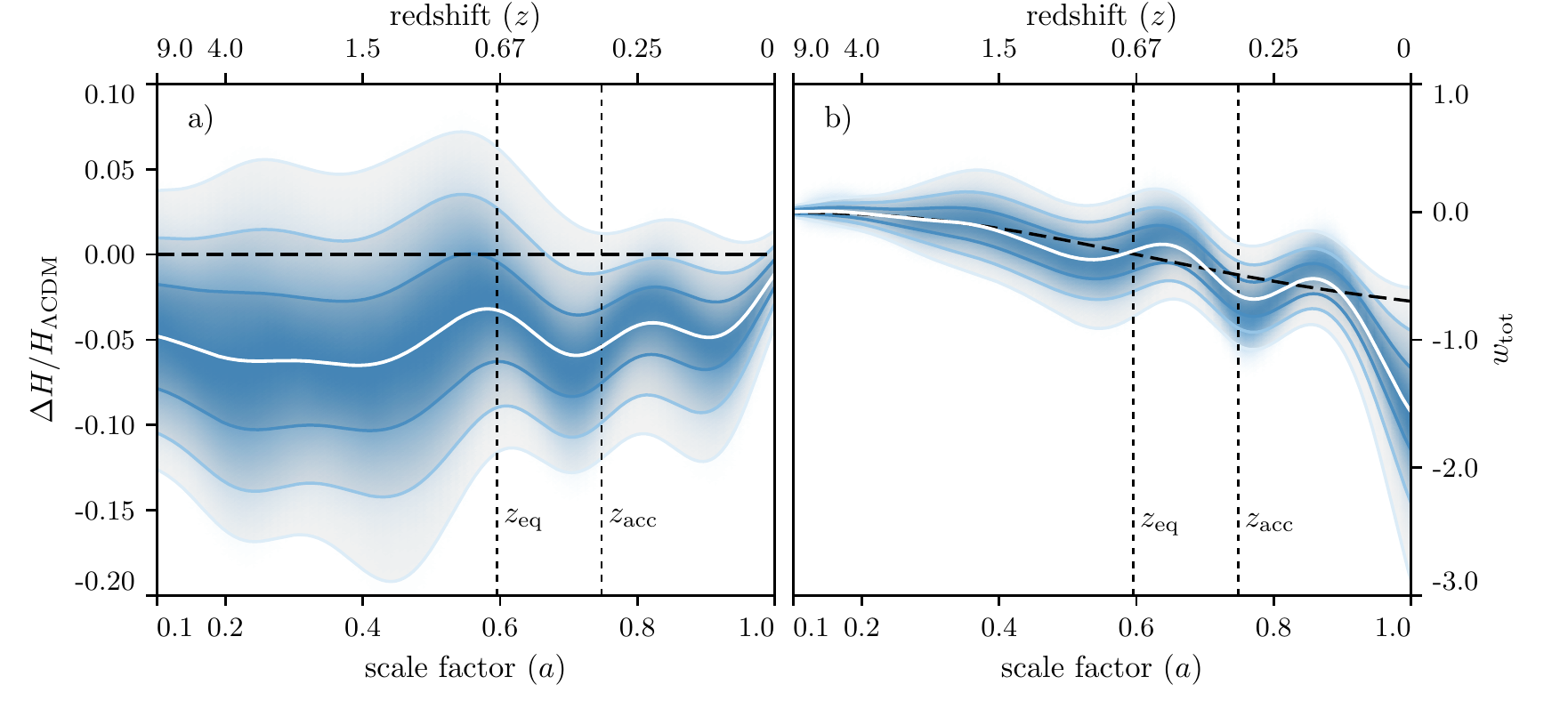}
\caption{ \label{fig:SHDerivedReconstruction} 
{\bf Reconstruction of Scalar Horndeski models.}
The marginalized distribution of variations in the expansion history, {\it Panel (a)}, and in the total effective equation of state, {\it Panel (b)}, in Scalar Horndeski models.
In both panels the white line represents the mean of the distribution while the other contours represent the $68\%$, $95\%$ and $99.7\%$ C.L. contours respectively.
The shade represents the probability distribution. The two dashed lines represent the redshift of equality between DM and DE and the redshift of the beginning of cosmic acceleration in the best fitting $\Lambda$CDM model.
}
\end{figure*} 
\begin{figure*}[!htbp]
\centering
\includegraphics[width=\textwidth]{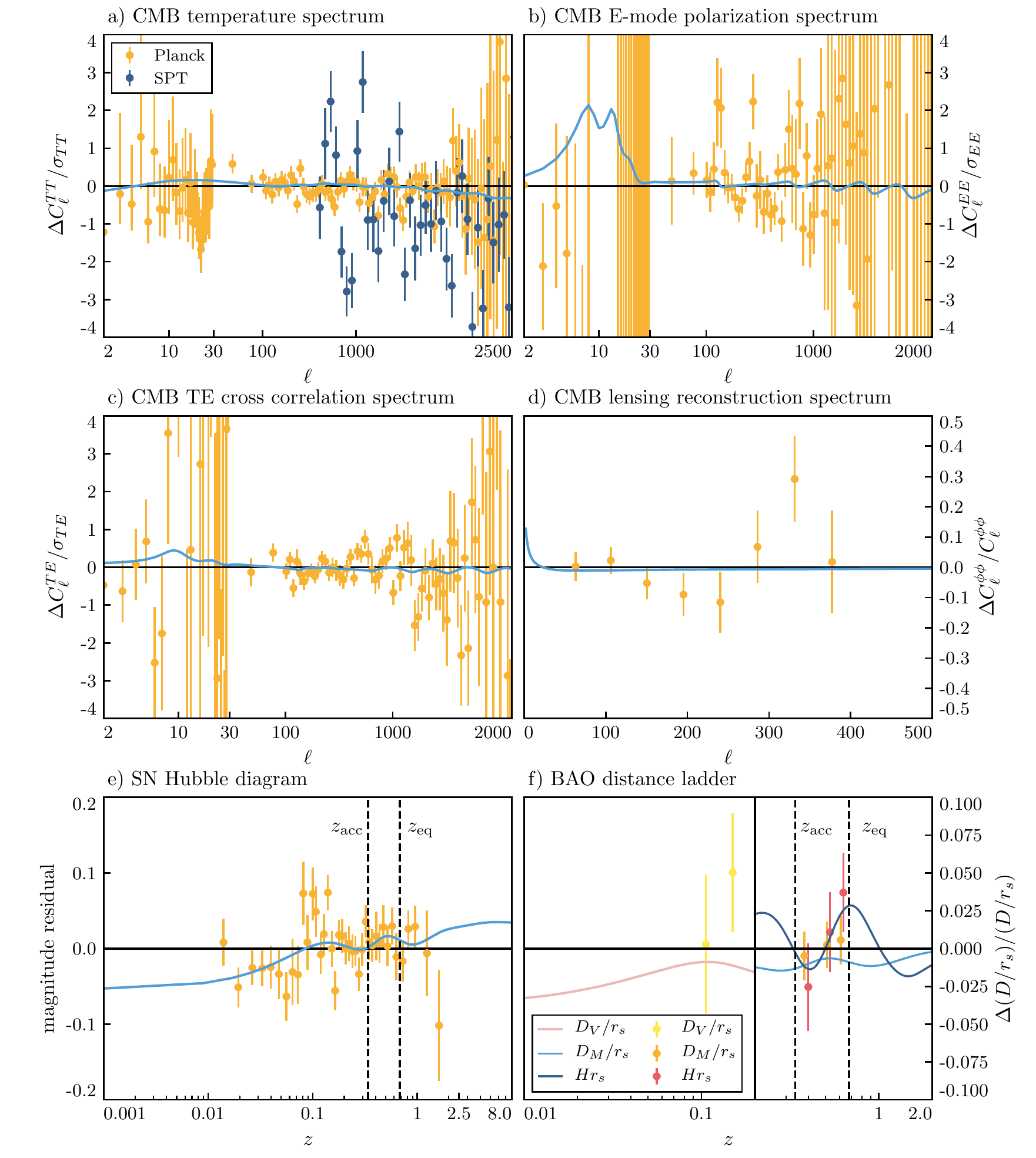}
\caption{ \label{fig:SHBestFitObservables} 
{\bf Reconstruction of Scalar Horndeski models.}
The best fit cosmological observables for Scalar Horndeski models compared to the the best fit $\Lambda$CDM cosmology.
{\it Panel (a)} shows the relative comparison of the CMB temperature power spectrum and {\it Planck} and SPT data points, in units of cosmic variance $\sigma_{TT}\equiv \sqrt{2/(2\ell +1)} C_\ell^{TT}$.
{\it Panels (b)} and {\it (c)} show the relative comparison of the CMB polarization power spectra, with {\it Planck} data points, in units of cosmic variance $\sigma_{EE}\equiv \sqrt{2/(2\ell +1)} C_\ell^{EE}$ and $\sigma_{TE} \equiv \sqrt{2/(2\ell +1)} \sqrt{ C_\ell^{TT}C_\ell^{EE} +(C_\ell^{TE})^2}$.
{\it Panel (d)} shows the relative comparison of the CMB reconstructed lensing potential and data points from {\it Planck}.
{\it Panel (e)} shows the magnitude residuals and the measurements from the Pantheon SN sample.
For better visualization we show the binned data points while the fitting procedure employs the complete catalog.
{\it Panel (f)} shows the relative comparison of BAO observables and their corresponding measurements. Notice that the three $D_M/r_s$ and $H r_s$ measurements are slightly displaced to help the reading of the figures but correspond to the same redshift.
}
\end{figure*} 

In this section we present the reconstruction of Scalar Horndeski (SH) models.
This is shown in Fig.~\ref{fig:SHReconstruction}\hyperref[fig:SHReconstruction]{a,c,e} in terms of the EFT functions $\Delta\Lambda/\Lambda$, $\Delta M_P^2/M_P^2$ and $\alpha_B$. 
The limit of this model to the $\Lambda$CDM one is represented by $\Delta\Lambda/\Lambda=0$, $\Delta M_P^2/M_P^2=0$ and $\alpha_B=0$ at all times.

With respect to the previous models the SH model family does not have significant differences in its phenomenology, beyond combining all the physical effects that we have seen for the KGB and GBD models.
We highlight that: $\Delta\Lambda/\Lambda$ influences the cosmological background and does not have a strong impact on the perturbations, as we discussed in Sec.~\ref{Sec:Quintessence}; $\Delta M_P^2/M_P^2$ influences both the background and the perturbations, as we discussed in Sec.~\ref{Sec:GBD}; $\alpha_B$ only influences the perturbations, as we discussed in Sec.~\ref{Sec:KGB}.
From Fig.~\ref{fig:SHReconstruction}\hyperref[fig:SHReconstruction]{a,c,e} we see that all three EFT functions do not appear to be significantly deviating from $\Lambda$CDM cosmology, even though the improvement in data fit is significant, as we have seen in Sec.~\ref{Sec:GeneralResults}.
As it happens for the other models that we discussed, this is an effect of poorly constrained functional modes that are being marginalized over a wide prior space.

We then compute the KL decomposition of the posterior and show it in Fig.~\ref{fig:SHReconstruction}\hyperref[fig:SHReconstruction]{b,d,f}.
For this model too we see that the number of constrained KL modes agrees with the estimate of the number of effective parameters, as in Tab.~\ref{Table:ReconstructionNeff}.
The number of modes that are being constrained for each function is, as we would expect, lower with respect to the models with less functions because their effect on cosmological observables is somewhat degenerate.
In particular we see only one mode being constrained for $\Delta\Lambda/\Lambda$ and not very strongly with respect to the prior.
This happens because of the marked degeneracy with $\Delta M_P^2/M_P^2$, that we have discussed in Sec.~\ref{Sec:GBD}.
On the other hand we see that we have plenty of constrained modes for $\Delta M_P^2/M_P^2$ but that its posterior is possibly contaminated by weekly constrained modes.
The same happens for $\alpha_B$. Several modes are constrained but the sum of the most constrained ones does not reproduce the best fit behavior.
Interestingly the overall constraining power of the data increases as the total number of constrained modes globally increases.
This reflects the fact that with increased model complexity the data is able to express stronger constraints.

We then project the reconstruction over the best constrained modes filtering out the poorly constrained ones.
This singles out the data features in the best fitting subspace, shown in Fig.~\ref{fig:SHFilteredReconstruction}.
As we expected from the KL decomposition we see that the overall trend is markedly different from the posterior.
In particular, as explained in the previous sections, the constraints become tighter, reflecting the fact that this figure is representing the minimal subspace that is needed by the model to provide a good fit to the data.

From this figure we can now see deviations that more accurately reflect the statistical significance of the reported deviations.
Cleaning out the most prior polluted modes, as it happened for other models, reveals some interesting features. 
In particular $\Delta\Lambda/\Lambda$ seems to transition around the redshift of equality.
On the other hand the behavior of $\Delta M_P^2/M_P^2$ does not seem to follow any of the two time scales.
Interestingly $\alpha_B$ exhibits an oscillation between the redshift of equality and the redshift of acceleration.

We have verified that this trend, and correspondingly good performances with the data, could not be achieved by simpler models because of stability requirements.
On their own or combined pairwise the behavior of these three EFT functions would not result in stable theories. 

Following the discussion in the previous sections we now turn to the projection of the constraints over model properties that we show in Fig.~\ref{fig:SHDerivedReconstruction}.
In Panel (\hyperref[fig:SHDerivedReconstruction]{a}) we see variations with respect to the $\Lambda$CDM expansion history while in Panel (\hyperref[fig:SHDerivedReconstruction]{b}) we show total effective equation of state.
With respect to KGB models we see that now the GR limit can be approached with negative variations of the expansion history.
With respect to GBD models we see that negative deviations are allowed to be much larger.
This is consistent with the fact that the GBD model is limited by stability boundaries in its performances. To reach lower values of the expansion history, with respect to $\Lambda$CDM the equation of state has to go further below $w=-1$ and hence requires stronger stabilization.
This is clearly shown in Panel (\hyperref[fig:SHDerivedReconstruction]{b}) that we can compare to Fig.~\ref{fig:GBDDerivedReconstruction}\hyperref[fig:GBDDerivedReconstruction]{b} to see that the equation of state can reach much smaller values.
Notice that in this model too, due to the inclusion of $\Delta M_P^2/M_P^2$ in the free EFT functions we cannot compute the DE equation of state as it becomes easily ill-defined when the effective DE density crosses zero. The expansion history is regular, as can be seen in Fig.~\ref{fig:SHDerivedReconstruction}\hyperref[fig:SHDerivedReconstruction]{a} and this is just a pathology of the effective definition. We then necessarily consider $w_{\rm tot}$.
The overall constraints on the expansion history are, in this model, at the level of about $20\%$ (at $99.7\%$ C.L.) around the redshift of equality.

Since the preference for the SH model is somewhat significant we now comment on the data details of the best fit cosmological solution.
We show the comparison of the best fitting SH model with respect to the best fit $\Lambda$CDM cosmology in Fig.~\ref{fig:SHBestFitObservables}.
The best fit value of the Hubble constant is $H_0=70.7$ and clearly improves the fit to local measurements.
As we can see from the figure, the fit to the CMB spectra is mostly the same as the $\Lambda$CDM model, compatible with the fact that these models are not solving the CMB lensing tension.
On the other hand we see that the fit to the SN and BAO data is improved. In particular the fit to the BOSS DR12 BAO measurements is improved but the solution is somewhat penalized by 6DF and MGS low redshift measurements.
SN in turn have a slightly improved fit as a result of a better fit to the low redshift end and the oscillations around the redshift of acceleration.
Notice that the oscillatory features in the data, that the model is capturing, seem to happen between the redshift of equality and the redshift of acceleration for both SN and BAO measurements.

We highlight that, in the best-fit model, part of the discrepancy in the measurement of $H_0$ is relieved by changing the low redshift end of the distance ladder between $z=0.01$ and $z=0.1$.
In the public likelihoods the $H_0$ measurement is implemented as a single constraint at redshift zero. 
This is, however, obtained by fitting the amplitude of the entire ladder, from the cepheids calibrators below $z<0.01$ to high redshift supernovae.
Most of the constraining power on the value of $H_0$ comes from redshifts below $z=0.1$ and the cosmological solution that we have found might be influenced by the inclusion of the lowest redshift measurements in the fitting pipeline.

The existence of this best-fit solution seems to contradict the conclusions of~\cite{Bernal:2016gxb,Aylor:2018drw} in the case where the spline reconstruction for the time dependence of the expansion history was used.
We highlight here that in this reconstruction the cosmological background is allowed to vary on a shorter time scale than the one considered in~\cite{Bernal:2016gxb,Aylor:2018drw} and, as such, the EFT models are allowed to provide a better fit to the SN and BAO data sets themselves, thus changing their inference.

%%%%%%%%%%%%%%%%%%%%%%%%%%%%%%%%%%%%%%%%%%%%%%%%%%%%%%%%%%%%%%%%%%%%%%%%
\subsection{Full Horndeski} \label{Sec:FH}
%%%%%%%%%%%%%%%%%%%%%%%%%%%%%%%%%%%%%%%%%%%%%%%%%%%%%%%%%%%%%%%%%%%%%%%%
%
\begin{figure*}[!htbp]
\centering
\includegraphics[width=\textwidth]{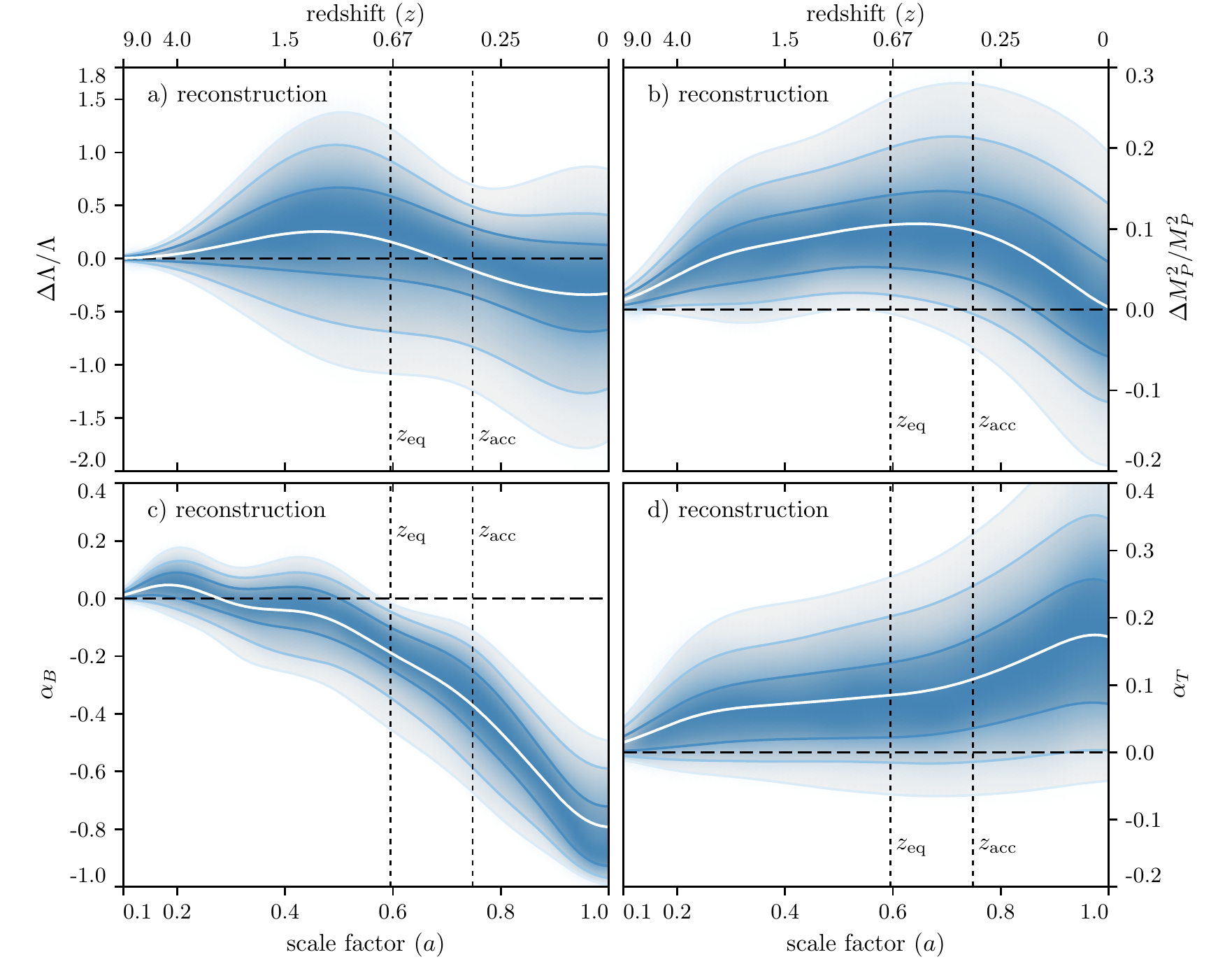}
\caption{ \label{fig:FHReconstruction} 
{\bf Reconstruction of Full Horndeski models.}
The marginalized posterior distribution of the EFT functions $\Delta \Lambda/\Lambda$, $\Delta M_P^2/M_P^2$, $\alpha_B$ and $\alpha_T$, describing all Horndeski models, as a function of scale factor and redshift.
The white line shows the mean of the distribution while the other contours represent the $68\%$, $95\%$ and $99.7\%$ C.L. regions respectively. The shade represents the posterior probability distribution. 
The two dashed lines show the redshift of equality between DM and DE, $z_{\rm eq}$, and the redshift of the beginning of cosmic acceleration $z_{\rm acc}$ in the best fitting $\Lambda$CDM model.
}
\end{figure*} 
\begin{figure*}[!htbp]
\centering
\includegraphics[width=\textwidth]{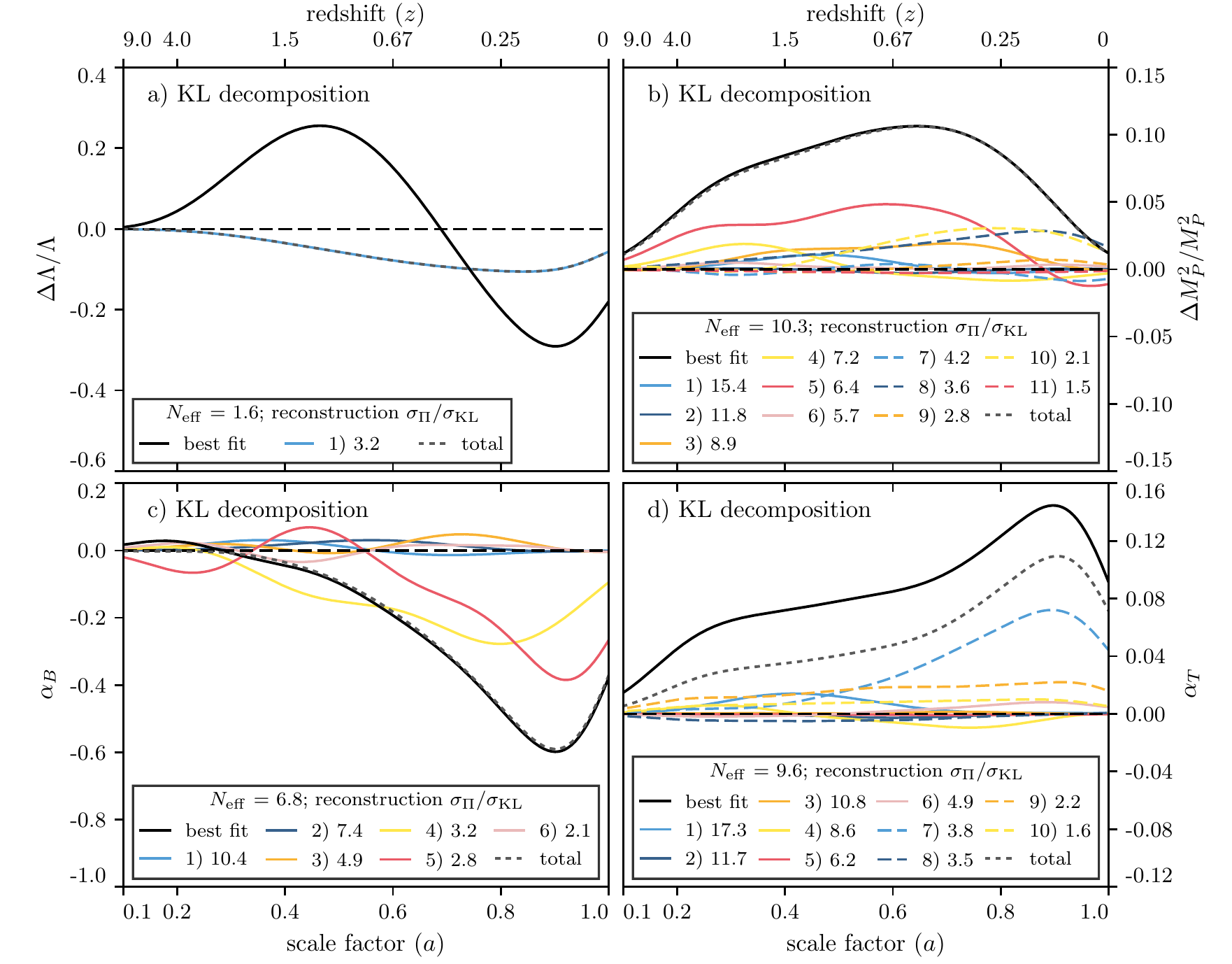}
\caption{ \label{fig:FHKLmodesReconstruction} 
{\bf Reconstruction of Full Horndeski models.}
The KL decomposition of the best fitting Horndeski cosmological model. 
The continuous black line represent the best fit model.
Different lines correspond to different KL modes, as shown in legend.
The black dashed line shows the model obtained as the sum of the KL modes that are shown.
}
\end{figure*} 
\begin{figure*}[!htbp]
\centering
\includegraphics[width=\textwidth]{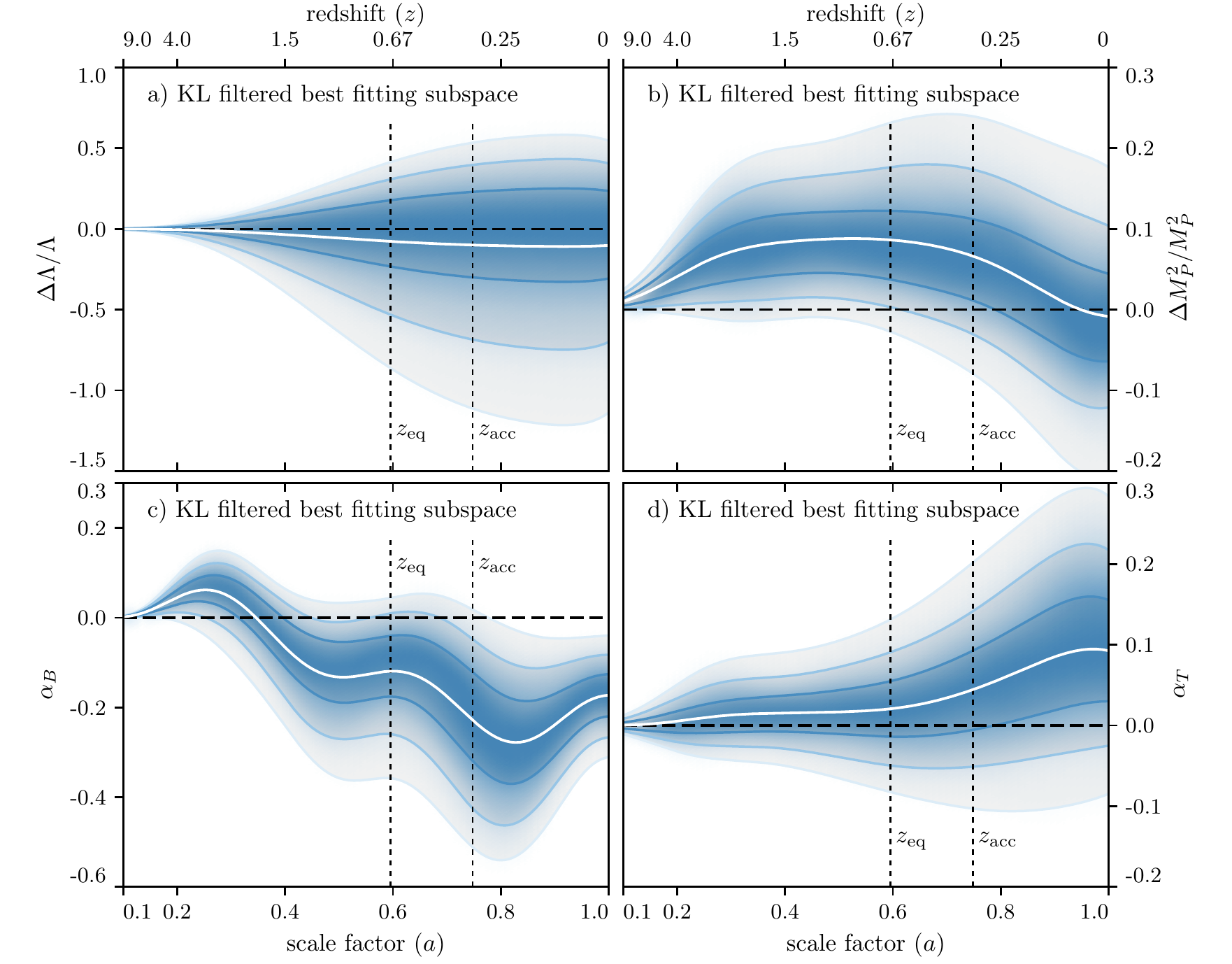}
\caption{ \label{fig:FHFilteredReconstruction} 
{\bf Reconstruction of Full Horndeski models.}
The marginalized distribution of the best fitting subspace for the reconstruction of Horndeski models.
The white line shows the mean of the distribution while the other contours represent the $68\%$, $95\%$ and $99.7\%$ C.L. regions respectively. The shade represents the posterior probability distribution. 
The two dashed lines show the redshift of equality between DM and DE, $z_{\rm eq}$, and the redshift of the beginning of cosmic acceleration $z_{\rm acc}$ in the best fitting $\Lambda$CDM model.
KL modes have been filtered by requiring that $\sigma_{\rm KL}<3 \sigma_{\Pi}$.
}
\end{figure*} 
\begin{figure*}[!htbp]
\centering
\includegraphics[width=\textwidth]{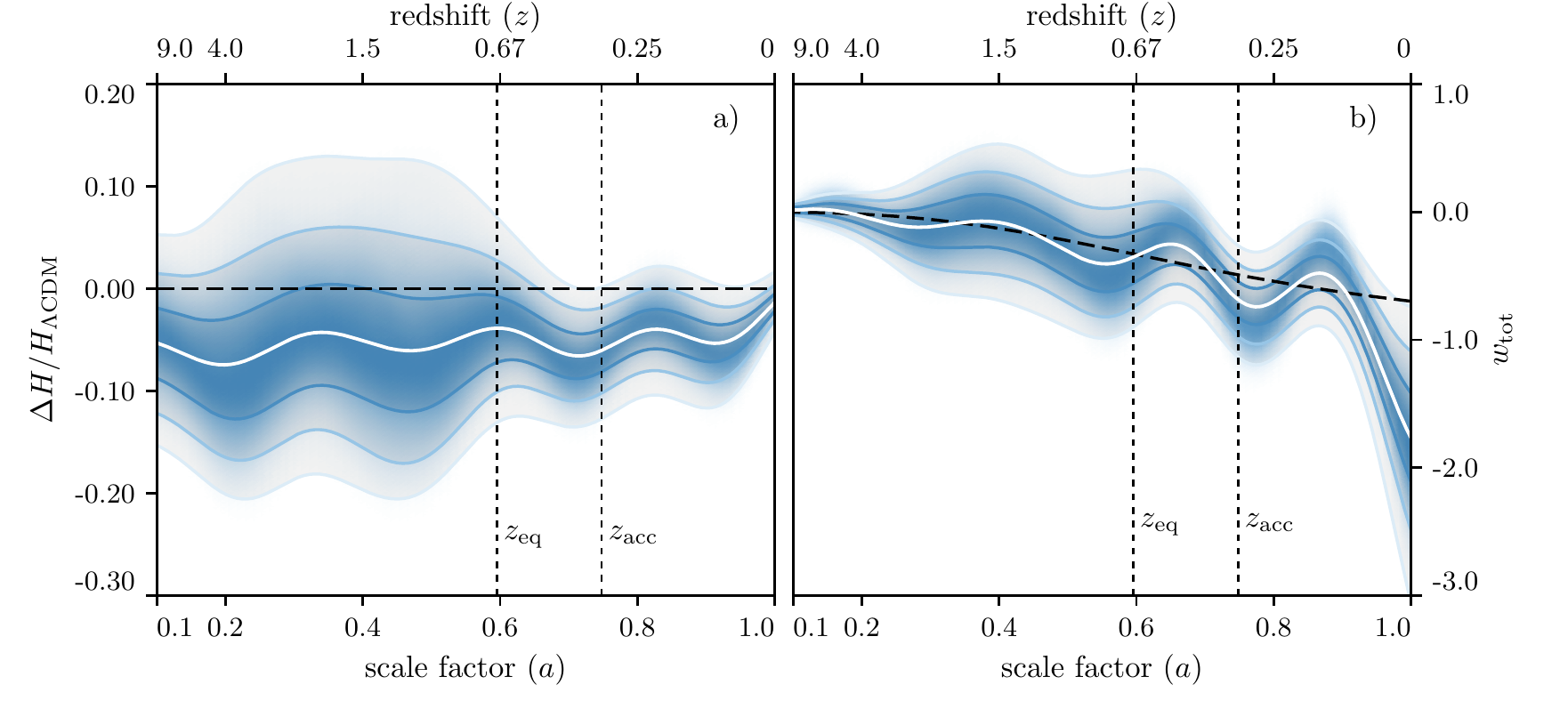}
\caption{ \label{fig:FHDerivedReconstruction} 
{\bf Reconstruction of Full Horndeski models.}
The marginalized distribution of variations in the expansion history, {\it Panel (a)}, and in the total effective equation of state, {\it Panel (b)}, in Horndeski models.
In both panels the white line represents the mean of the distribution while the other contours represent the $68\%$, $95\%$ and $99.7\%$ C.L. contours respectively.
The shade represents the probability distribution. The two dashed lines represent the redshift of equality between DM and DE and the redshift of the beginning of cosmic acceleration in the best fitting $\Lambda$CDM model.
}
\end{figure*} 

In this section we present the reconstruction of Full Horndeski (FH) models.
In this model all EFT functions are allowed to freely vary and the $\Lambda$CDM limit is given by $\Delta \Lambda/\Lambda=0$, $\Delta M_P^2/M_P^2=0$, $\alpha_B=0$ and $\alpha_T=0$.

The posterior of the reconstruction is shown in Fig.~\ref{fig:FHReconstruction} while its KL decomposition is shown in Fig.~\ref{fig:FHKLmodesReconstruction}.
Following the previous analysis the KL filtered best fitting subspace for the complete reconstruction is shown in Fig.~\ref{fig:FHFilteredReconstruction}.

The only novel aspect of this model, with respect to the model families that we have presented in the previous sections is the presence of $\alpha_T$ that modifies the speed of cosmological gravitational waves.
Notice that this is not the only effect that a change in $\alpha_T$ induces as it also modifies the behavior of the background and the perturbations.

As we can see this is very well constrained by cosmological observations and compatible with its GR limit. This is coherent with the fact that the best fitting cosmological solution for FH is very similar to that of SH.
With respect to the latter model, the constraints on the functions that they have in common is weakened, signaling a degeneracy with $\alpha_T$ but overall the number of modes that are being constrained by the data increases.

The similarity between the two models is also shown in the distribution of the derived quantities that we consider, as shown in Fig.~\ref{fig:FHDerivedReconstruction}, that does not present significant qualitative differences with respect to SH.

The cosmology and best fit behavior is not qualitatively changed with respect to the SH one, as shown in Fig.~\ref{fig:SHBestFitObservables}.

%%%%%%%%%%%%%%%%%%%%%%%%%%%%%%%%%%%%%%%%%%%%%%%%%%%%%%%%%%%%%%%%%%%%%%%%
\section{Conclusions} \label{Sec:Conclusions}
%%%%%%%%%%%%%%%%%%%%%%%%%%%%%%%%%%%%%%%%%%%%%%%%%%%%%%%%%%%%%%%%%%%%%%%%
%
In this paper we have presented the late-time reconstruction of DE and MG models on cosmological scales with current state of the art data.

We have found that some of the models, and in particular Scalar Horndeski (SH) and Full Horndeski (FH) models can, because of similar phenomenological features, relieve the Hubble constant discrepancy between CMB observations and local measurements.
As a consequence of the EFT treatment that we employed this resolution can be achieved in well defined DE or MG models that do not suffer from pathologies like ghost or gradient instabilities.
The best-fit solution leverages on the improved fit to SN and BAO measurements, and in particular on features in the expansion history that are compatible across these two data sets and seem to coincide with the redshift of the beginning of cosmic acceleration and the redshift of DM/DE equality.

These effects combined result in a preference for these two reconstructed models at a statistical level that is equivalent to $3.6\sigma$ if the phenomenological behavior that we identified is achieved with one extra parameter with respect to the $\Lambda$CDM model.

Noticeably these models do not address the discrepancies between the primary CMB spectra and the reconstruction of their lensing nor the tension between CMB measurements and the amplitude of perturbations as probed by weak lensing surveys.
Moreover, part of the tension on the value of $H_0$ is relieved at low redshifts and might be influenced by the approximate treatment of the lowest redshift end of the distance ladder as implemented in public likelihood codes.

Other simpler models, in particular Quintessence (Q), Kinetic Gravity Braiding (KGB) and Generalized Brans-Dicke (GBD), cannot achieve performances similar to SH or FH and are therefore strongly constrained.

We found that present cosmological probes are overall extremely sensitive to deviations from the $\Lambda$CDM model. This allows to constrain several DE/MG degrees of freedom, ranging from about $4$ in simple Q models to approximately $28$ for FH models.

% results model by model:
Model by model we have derived constraints on their characteristic functions and in particular:
variations in the cosmological constant $\Delta \Lambda/\Lambda$ are constrained, at $99.7\%$ C.L., at the $10\%$ level in Q models up to $150\%$ in FH, interestingly allowing for negative values of the cosmological constant;
variations in the Planck mass $\Delta M_P^2/M_P^2$ are constrained at the $20\%$ level in all models, at $99.7\%$ C.L.;
the kinetic term $\alpha_K$ is found to be unconstrained for all considered models;
kinetic braiding, $\alpha_B$, displays modes that are weekly constrained by the data while some of the constrained modes are an integral part of the resolution of the Hubble tension;
variations in the speed of gravitational waves, $\alpha_T$, are constrained at the $20\%$ level, at $99.7\%$ C.L., and largely compatible with GR.
% results on expansion history:
The constraints on the expansion history vary, depending on the flexibility of the model, and range from $4\%$ to $40\%$.

% future work:
The work toward the full exploitation of reconstruction techniques within the EFT of DE and MG is far from over.
Once the EFT characteristic functions have been obtained from the data, specific models and model properties can be derived. The projection of the reconstructions that we have presented to model space is an ongoing effort.
To see whether further discrepancies, and in particular the ones related to the lensing of the CMB and the primary CMB spectra can be solved we have to extend the reconstruction to cover models that display modification to the standard behavior at earlier times and in particular at the time of recombination.
Higher order operators or non-minimal couplings to different matter species can be added to these studies as well.
The practical feasibility of this program scales logaritmically with the number of operators that are considered so that their inclusion should be feasible in the near future.

%%%%%%%%%%%%%%%%%%%%%%%%%%%%%%%%%%%%%%%%%%%%%%%%%%%%%%%%%%%%%%%%%%%%%%%%
\acknowledgments
%%%%%%%%%%%%%%%%%%%%%%%%%%%%%%%%%%%%%%%%%%%%%%%%%%%%%%%%%%%%%%%%%%%%%%%%
%
We thank
Robert Crittenden,
Bin Hu,
Wayne Hu,
Bhuvnesh Jain,
Matteo Martinelli,
Samuel Passaglia,
Levon Pogosian,
Alessandra Silvestri,
Kimmy Wu,
and Gong-Bo Zhao
for useful comments.
MR is supported by U.S. Dept. of Energy contract DE-FG02-13ER41958.
Computing resources were provided by the University of Chicago Research Computing Center through the Kavli Institute for Cosmological Physics at the University of Chicago.

\clearpage

%%%%%%%%%%%%%%%%%%%%%%%%%%%%%%%%%%%%%%%%%%%%%%%%%%%%%%%%%%%%%%%%%%%%%%%%
\appendix
%%%%%%%%%%%%%%%%%%%%%%%%%%%%%%%%%%%%%%%%%%%%%%%%%%%%%%%%%%%%%%%%%%%%%%%%
%
%%%%%%%%%%%%%%%%%%%%%%%%%%%%%%%%%%%%%%%%%%%%%%%%%%%%%%%%%%%%%%%%%%%%%%%%
\section{Details of the EFT implementation} \label{App:EFT}
%%%%%%%%%%%%%%%%%%%%%%%%%%%%%%%%%%%%%%%%%%%%%%%%%%%%%%%%%%%%%%%%%%%%%%%%
%
In this appendix we discuss the details of the EFT parametrization that we use, with a particular focus on the notation and the strategy to set the cosmological background, mostly following~\cite{Raveri:2017qvt}.

We write the EFT action following~\cite{Gubitosi:2012hu,Bloomfield:2012ff} as the sum:
\begin{align}
S_{\rm EFT} \equiv S_{0,1} + S_2 +\dots + S_m \,,
\end{align}
where $S_{0,1}$ denotes the background and linear parts of the action, $S_2$ the quadratic part of the action and the dots represent higher order contributions to the action that are irrelevant on linear cosmological scales.
$S_{0,1}$ is the part of the action that influences the cosmological background while $S_2$ describes the behavior of linear perturbations.
$S_m$ is the matter action that contains all standard matter components (Dark Matter, baryons, radiation and neutrinos) that are assumed to be minimally coupled to gravity.
Hereafter the subscript $m$ indicates the sum over all standard matter species.

The background action, in the notation of~\cite{Hu:2014oga}, reads:
\begin{align}
S_{0,1} \equiv \int d^4x \sqrt{-g}  \bigg( &+\frac{M_{P0}^2}{2} \left[1+\Omega(\tau)\right] R \nonumber \\
& + \Lambda(\tau) - c(\tau)\,a^2\delta g^{00} \bigg) \,,
\end{align}
where $g\equiv \det (g_{\mu\nu})$, $M_{P0}^2 \equiv 1/(8 \pi G_N)$ is the present day value of the Planck mass, $R$ is the four-dimensional Ricci scalar, $a$ denotes the scale factor, $\tau$ is conformal time and $\delta g_{00}$ is the perturbation to the time-time component of the metric.
The background action is controlled by three arbitrary functions of time $\Omega$, $\Lambda$ and $c$.
In what follows, the accent mark represents a derivative with respect to the scale factor, the over-dot represents a derivative with respect to the conformal time.

The Friedmann equations resulting from this action are:
\begin{align}  
\label{Eq:FRW1}
\mathcal{H}^2 =&\, \frac{a^2}{3M_{P0}^2(1+\Omega)}(\rho_m+2c-\Lambda)-\mathcal{H}\frac{\dot{\Omega}}{1+\Omega} \,, \\
\dot{\mathcal{H}} =&\, -\frac{a^2}{6M_{P0}^2(1+\Omega)}(\rho_m+3P_m) -\frac{a^2(c+\Lambda)}{3M_{P0}^2(1+\Omega)} \nonumber \\
&\, -\frac{\ddot{\Omega}}{2(1+\Omega)} \,,
\label{Eq:FRW2}
\end{align}
where $\rho_m$ and $P_m$ denote the total matter energy density and pressure respectively and follow the standard matter continuity equations.

At the background level we then have four arbitrary functions of time, $\Omega$, $\Lambda$, $c$ and, as a consequence $\mathcal{H}$ with two constraint equations. 
This constraint is differential-algebraic and allows us to express two of these functions in terms of the other two.
That is, two of these four functions are assumed to be given and the other two functions are obtained by solving the constraint.

In the literature a common choice is to assume that $\Omega$ and $\mathcal{H}$ are given while $c$ and $\Lambda$ are computed through the constraint~\cite{Gubitosi:2012hu,Bloomfield:2012ff}. This is often called a designer approach since it allows to design an arbitrary expansion history in terms of, for example, an effective DE equation of state.

Following~\cite{Raveri:2017qvt} here we choose another basis and assume that $\Lambda$ and $\Omega$ are given as a function of scale factor and derive $c$ and $\mathcal{H}$ through the Friedman equations.

Once $\Lambda$ and $\Omega$ are given we can combine the two Friedman equations, convert to e-folds, $N\equiv \ln a$, introduce $y \equiv \mathcal{H}^2$, to obtain:
\begin{align}
\left( 1+\Omega+\frac{1}{2}a\Omega' \right)\frac{d y}{d N} &+\left(1+\Omega+2a\Omega' +a^2\Omega'' \right)y \nonumber \\
& +\left( \frac{P_ma^2}{M_{P0}^2} + \frac{\Lambda a^2}{M_{P0}^2}\right) = 0 \ ,
\label{Eq:hubble_app}
\end{align}
which is the linear, time-dependent, differential equation we solve to find $\mathcal{H}(a)$ once the EFT functions $\Omega$ and $\Lambda$ are given and with boundary condition $y(a=1)=\mathcal{H}_0^2$. 
Once this equation is solved the time dependence of $c$ can be obtained from the first Friedman equation, Eq.~(\ref{Eq:FRW1}), that can be written as:
\begin{align} \label{Eq:cConstraint}
\frac{ca^2}{M_{P0}^2} = \frac{3}{2}\left( 1+\Omega+a\Omega' \right)\mathcal{H}^2 -\frac{1}{2}\frac{a^2\rho_m}{M_{P0}^2} +\frac{1}{2}	\frac{\Lambda a^2}{M_{P0}^2} \,.
\end{align}
At this point the background is completely fixed and we can move to the discussion of the perturbations.
In this respect the second order action is not modified with respect to~\cite{Gubitosi:2012hu,Bloomfield:2012ff} and completely fixed once we fix the relevant EFT functions and the background evolution.

In this work we use the EFT basis of functions defined in~\cite{Bellini:2014fua} and the mapping to the EFT action that we use is simply given as in~\cite{Hu:2014oga}:
\begin{align} \label{Eq:GammaAlphaMapping}
\frac{M_P^2}{M_{P0}^2} =&\, 1+\Omega+ \gamma_3 \,, \nonumber \\
\alpha_K =&\, \frac{\frac{2ca^2}{M_{P0}^2}+4H_0^2\gamma_1a^2}{(1+\Omega+ \gamma_3)\mathcal{H}^2} \,, \nonumber \\
\alpha_B =&\,  +\frac{1}{2}\frac{a\gamma_2H_0+a\mathcal{H}\Omega^\prime}{\mathcal{H}(1+\Omega+ \gamma_3) } \,, \nonumber \\
\alpha_T =&\, -\frac{\gamma_3}{1+\Omega+ \gamma_3} \,,
\end{align}
in terms of the functions $\gamma_1$, $\gamma_2$ and $\gamma_3$ appearing in the second order action of ~\cite{Bloomfield:2012ff} in the convention of~\cite{Hu:2014oga}.
Notice that the convention for the definition of $\alpha_B$ is different from the one in~\cite{Bellini:2014fua} by a factor $-1/2$.
Eq.~(\ref{Eq:GammaAlphaMapping}) can be easily inverted to give:
\begin{align}
1+\Omega &= \frac{M_P^2}{M_{P0}^2}(1+\alpha_T) \,, \nonumber \\
\gamma_1 &= \frac{1}{4H_0^2a^2}\left[\alpha_K\frac{M_P^2}{M_{P0}^2}\mathcal{H}^2-\frac{2ca^2}{M_{P0}^2}\right]\,,  \nonumber \\
\gamma_2 &=  \frac{1}{aH_0}\left[+2\alpha_B\mathcal{H} \frac{M_P^2}{M_{P0}^2}-a\mathcal{H}\Omega^\prime\right] \,,  \nonumber \\
\gamma_3 &= -\alpha_T\frac{M_P^2}{M_{P0}^2} \,.
\end{align}
Notice that the dependence on $\Lambda$ in this mapping is effectively hidden in its dependence on the background expansion history and $c$.
In addition we can see that, with this change of EFT basis, the background expansion history depends also on both $M_P$ and $\alpha_T$ as we might expect would generically happen in models of DE/MG.

%%%%%%%%%%%%%%%%%%%%%%%%%%%%%%%%%%%%%%%%%%%%%%%%%%%%%%%%%%%%%%%%%%%%%%%%
\bibliography{biblio}
%%%%%%%%%%%%%%%%%%%%%%%%%%%%%%%%%%%%%%%%%%%%%%%%%%%%%%%%%%%%%%%%%%%%%%%%
\end{document}